\documentclass[reprint,
 amsmath,amssymb,
 aps, nofootinbib]{revtex4-2}

\usepackage{graphicx}
\usepackage{dcolumn}
\usepackage{bm}
\usepackage{hyperref}
\usepackage{physics}
\usepackage{xcolor}
\usepackage{float}
\usepackage{ulem}
\usepackage{subcaption}

\def\bb{\begin{eqnarray}}
\def\ee{\end{eqnarray}}
\newcommand{\kett}[1]{| #1 \rangle}
\newcommand{\braa}[1]{\langle #1 |}
\newcommand{\moy}[1]{\left\langle #1 \right\rangle}

\begin{document}

\title{A topologically protected quantum dynamo effect in a driven spin-boson model}

\author{Ephraim Bernhardt$^{1,*}$, Cyril Elouard$^{2,*}$, and Karyn Le Hur$^{1}$}
\affiliation{$^1$CPHT, CNRS, Ecole Polytechnique, Institut Polytechnique de Paris, Route de Saclay, 91128 Palaiseau, France}
\affiliation{$^2$ Inria, ENS Lyon, LIP, F-69342, Lyon Cedex 07, France}

\date{\today}

\begin{abstract}
We describe a quantum dynamo effect in a driven system coupled to a harmonic oscillator describing a cavity mode or to a collection of modes forming an Ohmic bosonic bath. When the system Hamiltonian changes in time, this induces a dynamical field in the bosonic modes having resonant frequencies with the driving velocity. This field opposes the change of the external driving field in a way reminiscent of Faraday's law of induction, justifying the term `quantum dynamo effect'. For the specific situation of a periodically driven spin-$\frac{1}{2}$ on the Bloch sphere, we show that the work done by rolling the spin from north to south pole can efficiently be converted into a coherent displacement of the resonant bosonic modes, the effect thus corresponds to a work-to-work conversion and allows to interpret this transmitted energy into the bath as work. We study this effect, its performance and limitations in detail for a driven spin-$\frac{1}{2}$ in the presence of a radial magnetic field addressing a relation with topological systems through the formation of an effective charge in the core of the sphere. We show that the dynamo effect is directly related to the dynamically measured topology of this spin-$\frac{1}{2}$ and thus in the adiabatic limit provides a topologically protected method to convert driving work into a coherent field in the reservoir. The quantum dynamo model is realizable in mesoscopic and atomic systems.
\end{abstract}

\maketitle
\def\thefootnote{*}\footnotetext{These authors have equally contributed to this work.}

\tableofcontents

\section{Introduction}\label{sec:introduction}

Understanding energy transfers at the nanoscale is of prime importance for the development of many technologies, requiring fundamental advances to many crucial tasks such as managing heat dissipation and heat-work conversion \cite{Sothmann14,Benenti17}, predicting the performances of quantum heat engines \cite{Uzdin15,Elouard17,Niedenzu18,Denzler20}, or quantifying and optimizing the resource cost to run manipulate quantum systems and run realistic quantum computers \cite{Bedingham16,Gea-Banacloche02,Auffeves22,Stevens21}.
Such topic lead to discover that the laws of thermodynamics, conceived centuries ago for macroscopic devices, actually hold at the nanoscale, and capture bounds on average energy flows received by a small classical \cite{Seifert12} or quantum system \cite{esposito2010} from external heat and work sources. This discovery, together with the growing ability to manipulate individual mesoscopic systems, certainly motivates the analysis of thermodynamics in the quantum regime, where new properties such as coherence \cite{Scully03,Hardal15,Streltsov17}, quantum measurement \cite{Elouard17,Ding18,Bresque21} or squeezing \cite{Manzano18,Niedenzu18} appear as additional resources.

Despite long-standing results in some regimes for the thermodynamics of weakly dissipative quantum systems \cite{spohn78,alicki79}, where the dynamics can be captured by the Gorini-Kossakowski-Sudarshan-Lindblad (GKLS) master equation \cite{cohentannoudjibook,breuerbook}, the definitions of heat and work still lack a unified description in the quantum regime.
 While there is a consensus to interpret energy changes associated with a time-dependent Hamiltonian as work performed onto the system (by an external classical field) \cite{alicki79,esposito2010}, it is an actively studied question to quantify work performed by a quantum system onto another \cite{Elouard22}. Besides, the energy provided by a reservoir initially at thermal equilibrium is usually considered as heat \cite{esposito2010,Potts21}, but recent studies showed it can contain contributions which exhibit work-like properties, e.g. in the case of coherent-dissipative systems \cite{monsel20,maffei21}. Studying all these phenomena requires to go beyond the reduced description of the system only and consider properties of the reservoirs or work sources it couples to. One approach to do so is to analyze microscopic models of the system and its environment, where the reservoir and its coupling to the system are described exactly via a joint system-reservoir Hamiltonian \cite{Nathan}. Famous examples are found in the widely studied class of Caldeira-Legget models \cite{caldeira1983quantum}, of which the Spin-Boson model is one instance, allowing for the demonstration of a large range of phenomena concerning e.g. quantum phase-transitions \cite{leggett1987dynamics, karynKLH, orth2013nonperturbative} or topology \cite{henriet2017topology}.

Here, we analyze the energy exchanges associated with the interaction of an externally driven quantum system with a bosonic environment which are initially at equilibrium at zero temperature. We show that under specific conditions, the system is able to displace the environmental modes, generating a coherent field in the reservoir.  This phenomenon, called ``Quantum dynamo" effect \cite{henriet2017topology}, can be understood as work-to-work conversion from the drive into the environment, mediated by dissipation. We first introduce general conditions to obtain a coherent field in the bath, demonstrating that the latter can be directly evaluated from the dynamics of the system observables coupled to the bath. We then focus on the case where the system is a spin-$\frac{1}{2}$ immersed in a rotating magnetic field to study this mechanism quantitatively. We show the special role of modes close to resonance with the driving frequency which carry the coherent displacement induced by the driving. We apply different approaches ranging from exact diagonalization, a numerically exact stochastic Schr\"odinger equation approach \cite{orth2013nonperturbative,henriet2017topology}, and the GKLS equation to simulate the dynamics of the spin and characterize the quantum dynamo effect in different regimes of the intensity of coupling to the reservoir and of the driving speed. We show that even if the phenomenon relies on disturbing the reservoir states, it survives in the weak coupling regime captured by the GKLS master equation. Increasing the coupling enhances the effect until an optimum is reached, after which too strong coupling leads to a vanishing dynamo effect as the spin state is frozen. Similarly, the dynamo power vanishes for too small drive velocities where dissipation effects become negligible. We take the point of view of thermodynamics to analyze the dynamo as an engine, demonstrating a trade-off between output power and efficiency as a function of the coupling strength and the driving speed. In the weak coupling limit, we capture the long-term behavior of the dynamo, showing that the efficiency reaches unity. Finally, we analyze the topological properties of the transformation performed on the spin through the implementation of a radial magnetic field, as captured by the Chern number and its dynamical counterpart \cite{henriet2017topology}, suggesting the dynamical protection of the dynamo energy transfer at long time, yet in the adiabatic driving limit. The driven spin-boson model exhibiting a quantum dynamo effect can be realized in circuit and cavity quantum electrodynamics \cite{Google2014,Boulder}.

The article is structured as follows. In Sec. \ref{sec:general}, we introduce general definitions related to the induced field in the bath, the dynamo effect and efficiency through a detailed analysis of different energies. In Sec. \ref{sec:one_mode}, we quantify the dynamo effect for one mode and discuss the importance of adjusting the driving velocity to the frequency of this specific mode.
Analytical solutions are derived for various frequencies of the mode in two special cases, namely on the one hand under the assumption of weak-coupling and adiabatic driving and on the other hand for the limit in which the spin remains polarized in the lab frame at all times. We present numerical results using exact diagonalization for a coupling to one mode and extended to a finite number of modes.
In Sec.~\ref{sec:multimode}, we discuss the situation of a periodically driven spin coupled to a continuous bath. We develop a GKLS Master equation approach to describe this setup in the long time limit. We compare to results from a numerically exact stochastic Schr\"odinger equation approach we employed and to a commonly applied approximation scheme (NIBA). We derive an analytical formula to quantify the induced field and its different contributions from the spin dynamics and consequently discuss the performance of the dynamo from the numerical results.
In Sec.~\ref{sec:topology}, for a half-period drive, we show a relation between the dynamo effect
and the dynamically measured Chern number through the energy transferred to the induced field. We also discuss the effect of a constant bias field.\\
In the appendices we present several derivations and numerical analyses to support our arguments of the main text.

\section{The quantum dynamo effect: setup and general definitions}\label{sec:general}

\begin{figure}
    \centering
    \includegraphics[width=\linewidth]{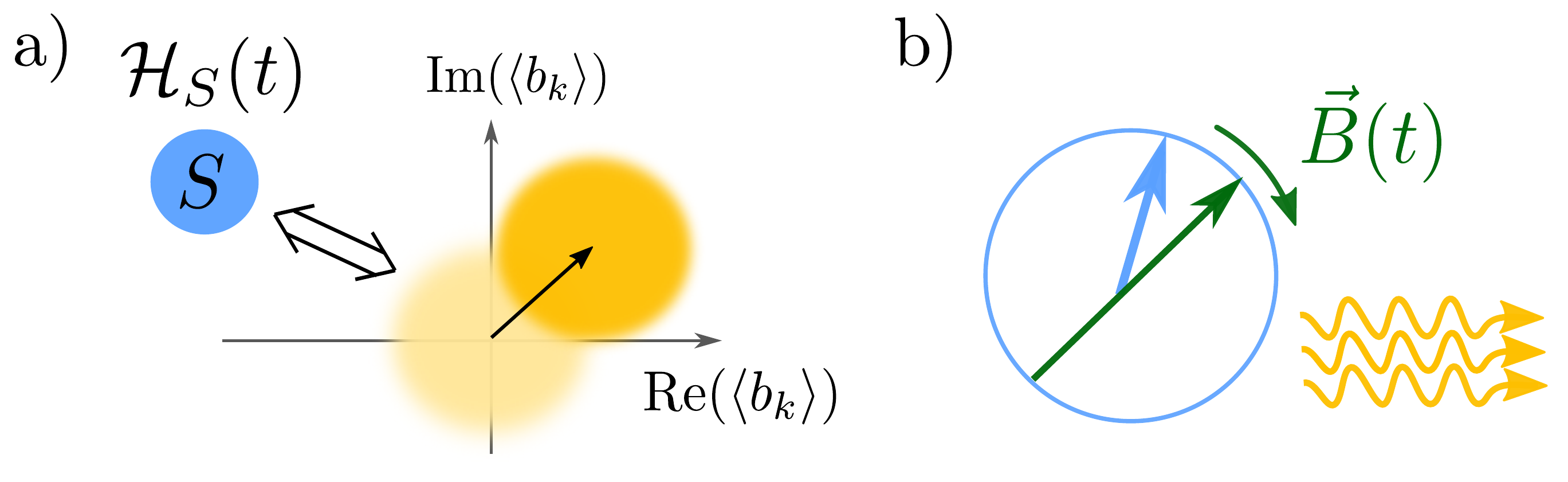}
    \caption{Principle of the quantum dynamo effect. {a) Coherently} driving a system coupled to a bosonic bath can induce a finite coherent displacement of some of the bath modes, i.e. a work-to-work conversion. b) As an application, we analyze the case of a spin (blue arrow) immersed in a magnetic field of fixed amplitude but rotating orientation (green arrow), which results in the emission of coherent light in the bosonic bath vacuum.}
    \label{fig:schematic}
\end{figure}

\subsection{Induced field}

In this section, we will set up the general definitions that we will use in subsequent sections to quantify the dynamo effect.
We introduce a system coupled to a bosonic bath, according to the Hamiltonian:
\bb
{\cal H} = {\cal H}_S(t) + SR + {\cal H}_R \label{d:basic_model}
\ee
where
\bb
{\cal H}_R &=& \sum_k \omega_k b_k^\dagger b_k,\nonumber\\
R &=& \sum_k g_k (b_k+b_k^\dagger).
\ee
The (free) Hamiltonian of the reservoir reads ${\cal H}_R$ with $b_k$ a boson operator, the reservoir observable coupled to the system is introduced through the symbol $R$ and $S$ is an observable of the system. Without lack of generality, we assume the coefficients $g_k$ to be real. The Planck constant $\hbar=\frac{h}{2\pi}$
is set to unity for simplicity.
The model \eqref{d:basic_model} we consider is a Caldeira-Leggett model \cite{caldeira1983quantum} in which the bath modes are non-interacting and linear. Note that in general, non-linearities can occur in many-body systems \cite{weissl2015kerr}, typically giving rise to finite lifetimes of the excitations \cite{samokhin1998lifetime, le2008charge}. However, model \eqref{d:basic_model} corresponds to different experimentally relevant situations, examples include superconducting circuits \cite{forn2017ultrastrong, magazzu2018probing} or a mesoscopic ring \cite{cedraschi2000zero, cedraschi2001quantum} coupled to an open transmission line, as well as a quantum dot coupled to a Chiral Luttinger liquid \cite{furusaki2002occupation, le2005unification}.

We first note that when the bath is at thermal equilibrium at inverse temperature $\beta$, namely $\rho_B^\text{eq} \propto e^{-\beta {\cal H}_B}$ (including the case of zero temperature), then the observable $R$ has a zero expectation value. This is related to the traditional view at weak coupling that a bath does not induce an average force on the system, but rather a fluctuating action which will be responsible for relaxation and decoherence, as captured e.g. by the GKLS equation.
However, as a result of the interaction with the system, the state of the bath is in general shifted from equilibrium.
With the particular form of the coupling Hamiltonian we consider $V = SR$, the average action of the system on the bath is to induce coherent displacements of the bosonic modes whenever $\moy{S(t)}$ is non-zero. This results in a non-zero expectation value of $R$ that we call in the following induced field $h(t) = \moy{R(t)}$, which can in turn generate an effective driving on the spin via the interaction Hamiltonian.
Such mean-field effect can be interpreted classically as the mechanical forces each system exerts on the other one, as it would also occur in the case of a bath of classical harmonic oscillators of position $x_k \propto \moy{b_k+b_k^\dagger}$ coupled to another classical system, say a classical rotor coupled to a piston at position $S\propto\cos(\phi)$, where $\phi$ is the rotor's rotation angle. In the full quantum case under study, quantum noise around these average forces leads to non-unitary reduced dynamics for the system and the bath, and contributes to the actual value of $\moy{R(t)}$ and $\moy{S(t)}$, as we discuss below.
In this Section, we provide an overview of when this phenomenon can occur in the situation of a driven system or time-dependent Hamiltonian ${\cal H}_S(t)$. The setup and a schematic summary of our study is shown in Fig. \ref{fig:schematic}.

The Heisenberg equation of motion for mode $k$ yields:
\bb
\label{eq:b_eom}
\moy{\dot b_k(t)} +i\omega_k \moy{b_k(t)}= -i g_k \moy{S(t)},
\ee
admitting the formal solution:
\bb
\label{eq:b_formal_solution}
\moy{b_k(t)} = e^{-i\omega_kt}\moy{b_k(0)} - i g_k\int_0^t dt'e^{-i\omega_k(t-t')}\moy{S(t')}.\nonumber\\
\ee
Assuming a zero initial expectation value of the modes and summing up the different modes' contributions we obtain an exact expression of the induced field:
\bb
h(t) &=& \int_0^t dt'K(t-t')\moy{S(t')}.\label{eq:ht}
\ee
in terms of the kernel function
\bb \label{eq:kernel}
K(t) = -2\sum_k g_k^2 \sin(\omega_k t),
\ee
that depends on the microscopic details of the bath. To get further intuitions, we will assume throughout the paper that the reservoir has an Ohmic spectral density, namely:
\bb \label{eq:ohmic_spectral_density}
J(\omega) = \pi\sum_k g_k^2\delta(\omega_k-\omega) = 2\pi\alpha\omega e^{-\omega/\omega_\text{c}},\ee
where $\omega_\text{c}$ is a characteristic cut-off frequency that we will consider to be the highest energy scale in the problem, and $\alpha$ is a dimensionless parameter characterizing the strength of the coupling. At frequencies $\omega\ll\omega_\text{c}$, the Ohmic spectral density is linear $J(\omega)\simeq 2\pi\alpha\omega$.
Using the $e^{-\omega/\omega_\text{c}}$ regularization of the spectral function in Eq. (\ref{eq:ohmic_spectral_density}), the kernel function simplifies to
\bb
K(t) = -\frac{8\alpha\omega_\text{c}^3 t}{(1+\omega_\text{c}^2 t^2)^2}.
\ee
 Other choices of the spectral density lead to different kernel functions, but qualitatively to a similar behavior. We briefly discuss the case of a hard cut-off at $\omega = \omega_\text{c}$ in Appendix~\ref{app:hard_cutoff}.

 Remarkably, the value of the induced field at time $t$ is entirely determined by the dynamics of the system observable $\moy{S(t)}$, and this result holds whatever the coupling strength, the nature of the system and of the coupled observables.
One can therefore exploit the large variety of analytical and numerical approaches giving access to the reduced system dynamics \cite{orth2013nonperturbative, henriet2017topology} to gain information about the bath state, see~Sec.~\ref{sec:multimode}.

 We can distinguish phenomenologically different cases leading to a non-zero induced field:
\begin{itemize}
    \item Case 1: $\moy{S(t)} = S_0$ is constant and non-zero (or converges at long times towards a non-zero constant): this may happen for instance when the system has a time-independent Hamiltonian and reaches equilibrium with the bath. In the case of the Ohmic bath we can perform the integration in Eq.~\eqref{eq:ht}. Focusing on long times $t\gg\omega_\text{c}^{-1}$, we obtain:
    \bb
        h(t) = -4\alpha\omega_\text{c} S_0,
    \ee
    which agrees with the equilibrium situation discussed above Eq. (\ref{eq:b_eom}).
    That is, a constant field opposed to $S_0$, which can be simply understood as the static displacement of the modes necessary to balance the constant force exerted by the system onto the bath. This force is typically zero if $S_0=0$. While this effect can be considered as a transfer of work to the bath, the energy exchange between system and bath stops when the equilibrium is reached. To obtain a sustained displacement of the bath, one conversely needs to consider time-dependent scenarios.
    \item Case 2:
    $\moy{S(t)}$ varies in time on a time-scale much longer than $\omega_\text{c}^{-1}$. This can occur because a parameter of the system's Hamiltonian is varied in time slowly.
    In this case, one can expect the induced field to follow adiabatically the value of $\moy{S(t)}$, i.e. for $t\gg\omega_\text{c}^{-1}$:
    \bb
    \label{eq:general_ad_h}
    h(t)=-4\alpha\omega_\text{c}\moy{S(t)},
    \ee
    Indeed, varying the value of $\moy{S(t)}$ ensures that energy exchanges keep occurring between the system and the bath. However, for a finite system the amount of energy that can be exchanged this way is bounded by the maximum value of $\moy{S(t)}$. Once reached, no more displacement can be induced in the bath. In other words, if one drives the system adiabatically along cycles $\moy{S(t+T)}=\moy{S(0)}$, the net displacement of the bath is zero.

    \item Case 3: $\moy{S(t)}$ varies in time in a non-adiabatic way. In this case it is possible to obtain additional dynamical contributions to the induced field, resulting from the interplay between the time-dependence of $\moy{S(t)}$ and the kernel $K(t)$. In the following, we denote (still $t\gg\omega_\text{c}^{-1}$):
    \bb
      h(t) &=& h_\text{ad}(t) + h_\text{dyn}(t)\label{d:hsplit}\\
      h_\text{ad} &=&-4\alpha\omega_\text{c}\moy{S(t)},\label{d:had}
    \ee
    to identify additional contributions with respect to the adiabatic case. In the present paper, we analyze such contributions, demonstrating that for a periodically driven system, it can be associated with a continuous energy flow to the bath responsible for a coherent displacement of its modes, that we call quantum dynamo effect.
\end{itemize}

\subsection{Dynamo energetics and efficiency}\label{sec:energetics}

We focus on the case of a periodically driven system, such that the quantum dynamo effect amounts to a conversion of work provided by the drive into energy in the form of coherent displacement of the bath modes.

We stress that this does not mean the reduced state of the bath is necessarily a Glauber coherent state.
Rather, due to the mean force exerted by the system on the bath, the Wigner function of each bosonic mode acquires a non-zero phase-space average $\moy{b_k}$, which is sufficient to induce an average force on the system.
To quantify this effect, we first introduce the energy stored in the bath in the form of coherent displacement, namely:
\bb
E_\text{dis}(t) = \sum_k \omega_k \vert\moy{b_k}\vert^2,
\ee
with $\moy{b_k}$ a function of time. We use again the Heisenberg equation for $b_k$ to express the associated energy flow:
\bb
\dot E_\text{dis}(t) =-\dot h(t) \moy{S(t)}.\label{d:dEdis}
\ee
In perfect agreement with the analogy to a classical oscillator bath mentioned above, the variation of $E_\text{dis}$ corresponds to the work of the average force induced by the system $F(t)\propto \moy{S(t)}$. In the quantum case, this can be formalized by checking that the rest of the bath energy change fulfills the second law of thermodynamics, which is detailed in Appendix~\ref{app:2ndLaw}, based on the results from \cite{Elouard22}.

On the other hand, the work provided by the drive can be evaluated as the energy change of the total quantum system composed of the bath and the system $S$, associated with the time-variation of the Hamiltonian ${\cal H}(t)$. This leads to:
\bb
\label{eq:drive_work_integral}
W_\text{dr}(t) = \int_0^t  dt'\moy{\frac{\partial {\cal H}(t')}{\partial t}}.
\ee

Finally, the quantum dynamo effect is a work-to-work conversion, associated with an efficiency $\eta_\text{dis} = \Delta E_\text{dis}(t)/W_\text{dr}(t)$, where and throughout the article, we denote $\Delta X(t) = X(t) - X(0)$ the variations of physical quantities with respect to the initial time of the transformation.
The emergence of work transfers between a quantum system and a reservoir initially at thermal equilibrium is a phenomenon that was recently emphasized in the case of a resonantly driven atom emitting photons into a waveguide \cite{monsel20,maffei21}.
The quantum dynamo effect that we analyze here constitutes a general mechanism for this phenomenon.\par
However, considering $E_\text{dis}$ to quantify the dynamo effect has two drawbacks: First, $E_\text{dis}(t)$ contains a contribution from the adiabatic part of the induced field which does not accumulate energy over many cycles of driving. From Eq. (\ref{eq:general_ad_h}), for the adiabatic component $E_\text{dis}(t)$ is proportional
to $\moy{S(t)}^2$. Assuming an adiabatic protocol, the variation of dissipated energy is zero for one period. Second, because of the spin bath interaction, the ground state of the joint system exhibits non-zero mode displacement depending on the value of $S$, which should be used as the proper reference for the displacements induced by the dynamo.
Otherwise, the non-zero initial interaction energy can be consumed to artificially increase the efficiency $\eta_\text{dis}$ above one (at least for short transformations), as discussed in Appendix~\ref{app:2ndLaw}.
In the ground state of the full system under assumption of quasistatic evolution, each mode $k$ has a displacement $\moy{b_k}=-(g_k/\omega_k)\moy{S(t)}$. We emphasize that in general in the ground state there is entanglement between the system and the bath. Since here we are examining the coherent displacement of the bath modes $\moy{b_k}$ in the ground state, we can employ the argument of shifting the oscillators depending on the system observable $\moy{S}$. \par
In the case where $\moy{S^2(t)}$ is a constant over the transformation, this issue can be addressed by defining the dynamo energy from the displacement of modes subtracting the adiabatic contribution, i.e. using it as a reference:
\bb
\label{eq:E_dyn}
E_\text{dyn}(t) = \sum_k \omega_k\vert\moy{b_k(t)}+\frac{g_k}{\omega_k}\moy{S(t)}\vert^2.
\ee
Its derivative is related to the non-adiabatic contribution of the induced field:
\bb
\label{eq:E_dyn_derivative}
\dot E_\text{dyn}(t) = h_\text{dyn}(t)\moy{\dot S(t)}.\label{d:dEdyn}
\ee
In addition, we quantify the average efficiency of the dynamo effect at a time $t$ via:
\bb\label{eq:efficiency}
\eta = \frac{\Delta E_\text{dyn}(t)}{W_\text{dr}(t)}.
\ee

Using that the spin and the reservoir undergo a unitary evolution ruled by Hamiltonian ${\cal H}(t)$, one can express the total energy balance in the form:
\bb
\label{eq:energy_balance}
W_\text{dr}(t) = \Delta E_{S} + \Delta E_\text{dyn} + \Delta E_\text{fluct}.
\ee
We have introduced $E_{S}(t) = \moy{{\cal H}_{S}(t)}$ and the energy stored in the fluctuation of the field around the dynamo displacement, namely:
\begin{widetext}
\bb
\label{eq:E_fluct}
E_\text{fluct}(t) &=& \sum_k \omega_k \bigg( \moy{ \left(b_k(t)^\dag + \frac{g_k}{\omega_k}S(t)\right)\left (b_k(t) + \frac{g_k}{\omega_k}S(t)\right)}- \left\lvert \moy{b_k(t)} + \frac{g_k}{\omega_k}\moy{S(t)} \right\rvert^2 \bigg),
\ee
\end{widetext}

which verifies $\Delta E_R(t) + \Delta E_\text{int} = \Delta E_\text{dyn}(t)+\Delta E_\text{fluct}(t)$, where $E_R(t) = \moy{\mathcal H_R(t)}$ is the bath energy and $E_\text{int}(t) = \moy{S(t)R(t)}$ is the energy stored in the coupling.
A detailed numerical analysis of the terms constituting this energy balance and of the contributions to $E_\text{fluct}$ can be found in Appendix \ref{app:fluctuations}. To write the energy balance, we have made the assumption $\moy{S(t)^2} = \text{cst}$. It can be understood in the following way: the work $W_\text{dr}(t)$ brought by the drive until time $t$ is shared by non-adiabatic excitations of the system $\Delta E_S(t)$, fluctuations and coherent displacement of the field $\Delta E_\text{fluct}(t)$ and $\Delta E_\text{dyn}(t)$. For the sake of simplicity, we will restrict the present analysis to initial states of the spin and bath verifying $E_\text{fluct}(0)=0$, which combined with the property that $E_\text{fluct}(t) >0$, leads to $\Delta E_\text{fluct}(t) >0$. Finally, noting $W_\text{dr}(t)\geq 0$ over one full rotation, we can quantify the efficiency of the dynamo setting $t=2\pi/v$:
\bb
\eta = 1- \frac{\Delta E_\text{fluct}(t)+\Delta E_S(t)}{W_\text{dr}(t)} \leq 1,
\ee
which implies that non-adiabatic excitation of the spin and the fluctuations of the field constitute sources of inefficiency for the quantum dynamo.\\

\subsection{Case of study: the rotated spin}\label{sec:case_of_study}

The exact conditions leading to the observation of the quantum dynamo effect, as well as its performances will depend on the actual implementation. Various parameters can play a significant role, as for example the speed of driving (we mentioned earlier that non-adiabaticity is required) or the strength of the coupling to the bath (non-zero dissipation is needed). Analyzing the role of such parameters requires to focus on an example. We do so in the remainder of this article by focusing on the situation where the quantum dynamo effect was first pinpointed \cite{henriet2017topology}, that is the case where the system $S$ is a single spin-$\frac{1}{2}$ subject to a magnetic field whose orientation is rotating. Such a situation is captured by the following Hamiltonian:

\bb
{\cal H} = {\cal H}_\text{spin}(t) + \frac{\sigma^z}{2}R+ {\cal H}_\text{bath},
\label{d:model}
\ee
where
\bb
{\cal H}_\text{spin}(t) = -\frac{H}{2}\left(\cos(vt)\sigma^z+\sin(vt)\sigma^x\right).
\label{d:system}
\ee
is the time-dependent Hamiltonian of the spin and we have identified $S = \sigma^z/2$, where $\sigma^i$, $i=x,y,z$ are the Pauli spin operators.

This spin Hamiltonian is a special instance of a spin in a radial magnetic field, which has topological properties.
The relation to a topological model constitutes a motivation for the choice of the system Hamiltonian \eqref{d:system} in this article.
We will explain how to define a dynamically measured Chern number and show how this is linked to the properties of the quantum dynamo effect in Sec. \ref{sec:topology}.

It is interesting for the following to introduce the frame rotating at frequency $v$ with respect to the lab frame, reached by the unitary transformation $U_\text{rot} = e^{-ivt\sigma_y/2}$. In this frame, the free dynamics of the spin (when the coupling to the bath is zero) is ruled by the time-independent Hamiltonian:
\bb
\label{eq:effective_spin_Hamiltonian}
{\cal H}_\text{spin}^\text{eff} = - \frac{H}{2}\sigma_z-\frac{v}{2}\sigma_y,
\ee
characterized by an effective spin energy splitting $\Omega = \sqrt{H^2+v^2}$ between its energy eigenstates.
One can then compute the unitary rotation induced by the drive on the spin in the absence of the bath:
\begin{widetext}
\bb
\moy{\sigma^x(t)} &=& \frac{H^2}{\Omega^2}\sin(vt)-\frac{v}{\Omega}\cos(v t)\sin(\Omega t)+\frac{v^2}{\Omega^2}\sin(v t)\cos(\Omega t),\label{eq:Usx}\\
\moy{\sigma^y(t)} &=& 2\frac{vH}{\Omega^2}\sin^2(\Omega t/2),\label{eq:Usy}\\
\moy{\sigma^z(t)} &=& \frac{H^2}{\Omega^2}\cos(vt)+\frac{v}{\Omega}\sin(v t)\sin(\Omega t)+\frac{v^2}{\Omega^2}\cos(v t)\cos(\Omega t).\label{eq:Usz}
\ee
\end{widetext}
One can see that in the case of an adiabatic (slow) rotation, i.e. $v\ll H$, the spin follows the orientation of the field, namely
\bb
\moy{\sigma^x(t)} \simeq \sin(v t)\\
\moy{\sigma^z(t)} \simeq \cos(v t)
\ee
At larger rotation speeds, non-adiabaticity manifests itself in the form of additional rotations of the spin state at frequency $\Omega$, which lead to non-vanishing values of $\moy{\sigma^y(t)}$.

\subsection{Effects of the bath on the spin}

The bath can have different effects on the spin depending on the intensity of the coupling and in relation with the driving.

For weak enough coupling, requiring in particular $\alpha\ll 1, H\ll \omega_\text{c}$, one can expect to capture the dissipation and dephasing induced by the bath via a Markovian quantum master equation for the spin.
We note that the spin Hamiltonian does not contain a drift (time-independent term) larger than the time-dependent part, which departs from the most common situation of weak driving. It must therefore be expected that the properties of the driving will strongly influence the characteristics of the dissipation.
At larger coupling, the entanglement between the spin and the bath grows, leading to an important phenomenon pointed out in \cite{leggett1987dynamics,orth2013nonperturbative}, for a fixed field angle (i.e. $vt \to \theta=cst$), which is the effective reduction of the spin tunneling coefficient (namely the Hamiltonian coefficient along $\sigma^x$). This phenomenon is associated with the growing spin-bath entanglement when $\alpha$ increases \cite{kopp2007universal}, which results in a reduced overlap between the instantaneous eigenstates associated with the spin pointing upward and upward. This effect becomes negligible for small values of $\alpha\ll 1/\log(\omega_\text{c}/\Delta)$.\par

Finally, the existence of a non-zero induced field (both adiabatic and non-adiabatic contributions) can also affect the dynamics of the field by modifying the unitary part of the spin dynamics. Before investigating these effects and their impact on the dynamo effect performances, it is useful to consider the case where the spin is coupled to only one mode. This special case will give us insights on the role of the different frequency ranges present in the bath spectrum and the mechanism at the core of the quantum dynamo.

\section{The quantum dynamo for one mode}\label{sec:one_mode}

In Ref.~\cite{henriet2017topology}, the quantum dynamo is exemplified on a model of a driven spin coupled to a single bosonic mode resonant with the driving velocity. In this Section, we revisit and extend these results. To be precise, we are in this Section considering the model of Eq. ~\eqref{d:model} with only one bosonic mode with a frequency $\omega$ such that the Hamiltonian reads
\begin{multline}
    \label{eq:single_mode_hamiltonian}
    \mathcal{H} = -\frac{H}{2}(\sin(vt) \sigma_x + \cos(vt) \sigma_z)
    + g (b + b^\dag)\frac{\sigma_z}{2} + \omega b^\dag b.
\end{multline}
Defining the induced field of the bosonic mode as
\begin{equation}
    h_{\omega}(t) = g \langle b+b^\dag \rangle,
\end{equation}
it fulfills
\begin{equation}\label{eq:one_mode_ODE}
  \frac{1}{\omega^2}\ddot{h}_\omega+h_\omega = -\frac{g^2}{\omega} \langle \sigma^z \rangle.
\end{equation}
We denote it with a subscript $\omega$, since this is the field induced onto the spin by one bosonic mode with a frequency $\omega$. The resonant frequency evoked in \cite{henriet2017topology} corresponds to $\omega = v$.

In order to analyze the dynamo effect more precisely, we first need to specify the protocol we consider, and in particular the initial spin-bath preparation. In the following we distinguish two cases:
\begin{enumerate}
    \item Preparation (1): The spin and the mode are initially in the ground state of Hamiltonian ${\cal H}(t)$ in Eq.~\eqref{eq:single_mode_hamiltonian}, at time $t=0$. A possible protocol to initialize such a state would be to apply a constant large magnetic field along the $z$-direction and let the coupled systems relax in contact with a zero-temperature environment. This ground state is a factorized spin-mode state, the spin being in the eigenstate of $\sigma_z$ pointing upward, and the mode being in the ground state of:
    \begin{equation}
    \mathcal{H}_\text{boson} = \frac{g}{2}(b+b^\dag) + \omega b^\dag b,
    \end{equation}
    that is the displaced vacuum veriyfing $b\ket{GS} = -g/(2\omega) \ket{GS}$. Finally, the initial condition associated with such preparation is
\begin{equation}
    h_{\omega}(0) = g(\langle b(0) \rangle + \langle b^\dag(0) \rangle) =  -\frac{g^2}{\omega}.
\end{equation}
    \item Preparation (2): The spin and the bosonic mode are initially not interacting and the spin is initialized in the state $\ket{\uparrow}_z$, the bosonic mode being unoccupied. At $t=0$, the interaction and the driving of the spin are turned on. This implies the initial condition $h_\omega(0)=0$.
\end{enumerate}

We can then solve Eq.~\eqref{eq:one_mode_ODE} to obtain:
\bb
h_\omega(t) &=& -g^2\int_0^t dt' \sin(\omega(t-t'))\moy{\sigma^z(t')}\nonumber\\
&&-\delta_{(1)} \frac{g^2}{\omega}\cos(\omega t). \label{eq:ht1}
\ee
where $\delta_{(1)}=1$ for preparation (1) and $0$ for preparation (2). Note that the second term corresponds to the free part of the induced field that one would expect without the influence of the spin dynamics. In the first term, we retrieve Eq.~\eqref{eq:ht} with a kernel function for one mode $K_1(t)=-2\sin(\omega t)$. The ratio of $\omega$ and the typical frequency scales ruling the evolution of $\moy{\sigma^z(t)}$ will determine whether an induced field can build up. Here we distinguish different regimes.

\subsection{Weak coupling and adiabatic driving}\label{sec:one_mode_adiabatic}

If the coupling to the mode is weak $g\ll H$, the spin evolution is expected to be close to the unitary evolution given by Eqs.~\eqref{eq:Usx}-\eqref{eq:Usz}. If in addition the rotation speed is small $v\ll H$, one can make the approximation $\moy{\sigma^z(t)} \simeq \cos(vt)$ in Eq.~\eqref{eq:ht1}, leading to:
\begin{equation}
  \label{h_onemode_cos_1}
  h_{\omega} = \frac{ g^2 \omega}{\omega^2-v^2} \left( \frac{v^2}{\omega^2}\cos(\omega t) - \cos(vt) \right),
\end{equation}
for preparation (1), and
\begin{equation}
  \label{h_onemode_cos_2}
  h_{\omega} = \frac{g^2 \omega}{\omega^2-v^2} \left( \cos(\omega t) - \cos(vt) \right),
\end{equation}
for preparation (2).

Then, the behavior strongly depends on the ratio between the mode frequency $\omega$ and the velocity $v$:
\begin{itemize}
    \item If $\omega\ll v$, the induced field simplifies to:
    \bb
    h_\omega(t)= -\frac{g^2}{\omega}\cos(\omega t) + {\cal O}\left(\frac{\omega^2}{v^2}\right),
    \ee
    for preparation (1), that is, the initial field evolves almost freely. For preparation (2), the induced field scales as $\omega^2/v^2$:
    \bb \label{eq:one_mode_v_gg_w}
    h_\omega(t) = -\frac{g^2}{\omega}\frac{\omega^2}{v^2}(\cos(\omega t)-\cos(v t))+ \frac{1}{v}{\cal O}\left(\frac{\omega^3}{v^3}\right).
    \ee

    \item If $\omega \gg v$, the mode sees a spin evolving quasi-statically and follows adiabatically the field  on average (with additional oscillations for the case of preparation (2)), i.e.:

    \bb\label{eq:one_mode_w_gg_v}
    h_\omega(t) &=&   -\frac{g^2}{\omega}(\cos(v t)-(1-\delta_{(1)})\cos(\omega t)) \nonumber\\
    &&+\frac{1}{v}{\cal O}\left(\frac{v^3}{\omega^3}\right).
    \ee

    \item For $\omega = v$, we finally get:
    \bb\label{one_mode_h_ind}
    h_v(t) &=& -\tfrac{1}{2}g^2t\sin(vt)-\delta_{(1)}\frac{g^2}{v}\cos(vt).
    \ee
\end{itemize}

One can use these expressions, the form of the adiabatic induced field in this limit $h_\text{ad}(t) = -\frac{g^2}{\omega}\cos(vt)$, and Eq.~\eqref{d:dEdyn}  to get a first intuition of the mechanism leading to the dynamo effect: modes far from resonance will accumulate no or little dynamo energy. Indeed, the non-adiabatic induced field for both slow and fast modes lead to either field contributions that have a frequency very different from $\moy{\dot\sigma^z(t)} \simeq -v\sin(vt)$ or adiabatic contributions $\propto \cos(vt)$. Both yield negligible energy variation $\Delta E_{\omega,\text{dyn}}$ over a cycle $t\in[0,2\pi/v]$. Conversely, a mode (quasi-)resonant with the driving frequency $\omega\simeq v$ will lead to a field of growing amplitude, and therefore to an accumulation of dynamo energy. It can be quantified using Eq. \eqref{eq:E_dyn_derivative} for one mode under the approximation $\langle \sigma^z(t) \rangle \sim \cos(vt)$ by
\bb
   \Delta E_\text{dyn} &=& \frac{g^2}{32 v} \bigg((4 \delta_{(1)}-3)(1-\cos (2 v t))\nonumber \\
   &+& 2 v^2 t^2 -2 v t \sin (2 v t)\bigg) \label{eq:one_mode_E_dyn}
\ee
Conversely, the work done to the system by driving defined in Eq. \eqref{eq:drive_work_integral} can in this weak coupling approximation be approximated by
\bb
    W_\text{dr}(t) &=& \frac{g^2}{32 v} \bigg((1+4\delta_{(1)})(1-\cos (2 v t)) \nonumber \\ &+& 2 v^2t^2 -2 vt \sin (2  v t)\bigg)
    \label{eq:one_mode_work}
\ee
The energy of fluctuations can be determined by
\begin{equation}\label{eq:one_mode_E_fluct}
    \Delta E_\text{fluct} = \frac{g^2}{4v} \sin^2(v t).
\end{equation}
From this consideration it is clear that here the work is always bigger than the coherent energy change of the displaced mode due to the dynamo effect. Both of them have contributions growing quadratically in time. In contrast, the energy of the fluctuations is periodic with constant period, so it is clear that the portion of work going into useful energy is increasing, which enhances the efficiency at long times $vt\gg 1$ where the ratio $\Delta E_\text{dyn}/W_\text{dr}$ goes to unity (if the approximation of nearly free and adiabatic spin dynamics remains true).

Note that from Eq. \eqref{one_mode_h_ind}, the situation for a half-period driving needs to be discussed with care since the term $-\frac{1}{2}g^2 t \sin(v t)$ is zero at initial $t=0$ and final $t=\frac{\pi}{v}$ times such that for preparation $(1)$, $\Delta h_v = h_v(\frac{\pi}{v})-h_v(0)= \frac{2 g^2}{v}$ whereas
for preparation $(2)$, $\Delta h_v = 0$ within this protocol. Then, we identify the dynamo effect from the definition of the averaged produced field
\begin{equation}
\int_0^{\pi/v} h_v(t')dt' = -\frac{1}{2} \frac{g^2}{v^2}\pi,
\end{equation}
for both preparations.
This corresponds to the dynamically induced field exerting a back-action on the spin. We also observe from Eq. \eqref{one_mode_h_ind} that the dynamically induced field opposes the effect of the external driving field with its amplitude growing linearly in time. This opposition to the change in the driving field is reminiscent of Faraday's law of induction, which motivated the term `quantum dynamo effect' in \cite{henriet2017topology}.

For the protocol studied in this section, the dissipated energy into the bath $\Delta E_\text{dis}(t=\pi/v)=-\int_0^{\pi/v} \dot{h}\langle S \rangle dt = \frac{g^2}{16}\frac{\pi^2}{v}$ is equivalent to the dynamo energy $\Delta E_{v,\text{dyn}}$ in Eq. \ref{eq:one_mode_E_dyn}
since the variation of interaction energy $h\langle S\rangle$ is zero between the two poles. Then, we find for the dynamo energy after every half period of the driving (i.e. at times $t = n\pi/v$ with $n \in \mathbb{Z}$
\bb
\label{eq:E_dyn_half_analytical}
\Delta E_{v,\text{dyn}}(t=n\pi/v) = \frac{n^2 g^2 \pi^2}{16 v}.
\ee
We note that the dynamo energy scales as the square of the coupling strength $g$, such that going to larger coupling seems to allow for a powerful dynamo.
However, this result relies on the unitary approximation of the spin dynamics, and therefore remains valid only as long as $- h(t) \ll H$, which according to Eq.~\eqref{one_mode_h_ind} requires $t \ll 2 H/g^2$.
Therefore, increasing $g$ too much will lead to a breakdown of the effect as we explain in next Section.

\subsection{Polarized spin limit}\label{sec:one_mode_polarized}

In presence of a strong induced field, the rotation of the spin might be frozen. Indeed, the spin dynamics can be considered to be ruled by an effective Hamiltonian ${\cal H}_\text{spin,eff} = (-H\cos(vt)/2+h(t)/2)\sigma^z -(H\sin(vt)/2)\sigma^x$. For $-h(t)>H\gg v$, the unitary evolution induced by such Hamiltonian remains in the northern hemisphere of the Bloch sphere (i.e. $\moy{\sigma^z(t)}>0)$). When $-h(t)\gg H$, the spin dynamics is almost completely frozen $\moy{\sigma^z(t)}\simeq 1$. Injecting this behavior into the induced field integral equation \eqref{eq:ht1}, we obtain:
\bb \label{eq:frozen_field}
h_\omega(t) = -\frac{g^2}{\omega}\left[1+(\delta_{(1)}-1)\cos(\omega t)\right].
\ee
We see that for preparation $(1)$, this behavior will be ensured by the condition $g^2/\omega \gg H$, whatever the driving speed, leading to a constant induced field and therefore the absence of the quantum dynamo effect.

The same mechanism leads us to also expect a breakdown of the quantum dynamo effect for coupling to only one mode after a long time of operation. Indeed, assuming that we are initially in the weak coupling regime and the mode is resonant $\omega =v$, we see from Eq.~\eqref{one_mode_h_ind} that the amplitude of the field grows as $h_v\sim g^2 t/2$, and becomes larger than the driving field $H$ for $t>H/2g^2$.

\subsection{Numerical results}

\begin{figure*}
    \begin{center}
    \includegraphics[width=\textwidth]{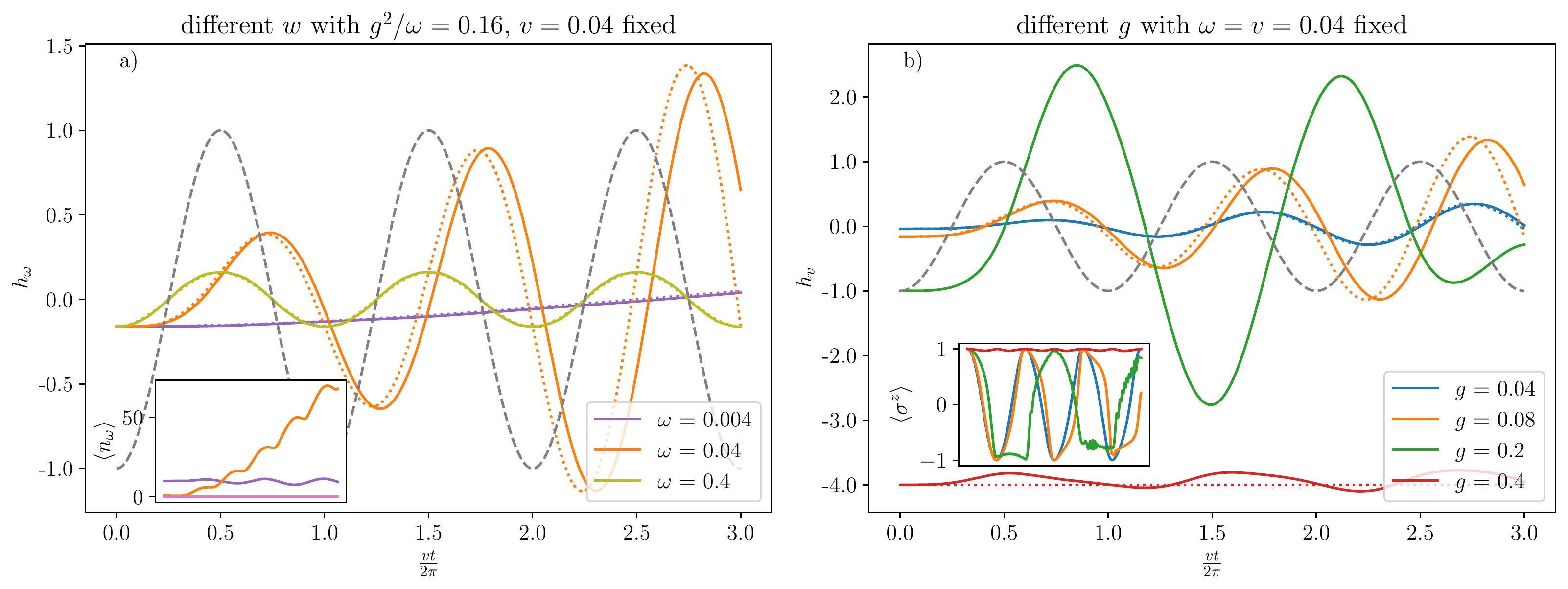}
    \end{center}
    \caption{One-mode dynamo. a) induced field as a function of $v t$ from numerical exact diagonalization of the Hamiltonian in Eq. \eqref{eq:single_mode_hamiltonian} for a total of three driving periods. Varying $\omega$ while keeping $g^2/\omega$ fixed, we observe a large induced field with an amplitude growing in time for the resonant mode (orange curve). The dotted lines show the comparison to Eqs. \eqref{eq:one_mode_v_gg_w}-\eqref{one_mode_h_ind} obtained under the approximation $\moy{\sigma^z} \approx \cos(vt)$. In the inset, we show the corresponding expectation values of the occupation number $\langle n_\omega \rangle$. b) Induced field of a resonant mode with $\omega = v$ for different values of the coupling $g$.
    The corresponding expectation value of $\moy{\sigma^z}$ is shown in the inset.
    The dotted lines show theoretical predictions from the unitary spin limit in Eq. \eqref{one_mode_h_ind} for the blue and the orange curve and from the frozen spin limit in Eq. \eqref{eq:frozen_field} for the red curve.
    For comparison, in both plots the external field in $z$-direction (i.e. $-H \cos(vt)$) is shown by a gray dashed line. In all simulations, $H=1.0$ and the system is initialized in preparation (1). The Hilbert space of the bosonic mode was truncated at an occupation of $N_b = 200$ and we checked that the achieved occupations are well below this number.
    }
    \label{fig:one_mode}
\end{figure*}

To confirm these conclusions, we solve the coupled evolution of the spin and one mode numerically using exact diagonalization (ED). In Fig. \ref{fig:one_mode} a), we show simulation results for different frequencies $\omega$ of the bosonic mode and starting from preparation (1).

The dynamo effect occurs at the resonance frequency $\omega = v$, where the induced field opposes the effect of the external driving field as predicted by Eq. \eqref{one_mode_h_ind}. This is similar to \cite{henriet2017topology}, where the dynamo effect has been identified from the induced field opposing the external driving field after driving for half a period, i.e. $t=\pi/v$. The resonant mode furthermore reaches a large occupation (shown in the inset of Fig. \ref{fig:one_mode} a)).
We examine the resonant case with fixed $\omega = v =0.04$ (in units of $H=1$) more closely in Fig. \ref{fig:one_mode} b) by plotting results for different coupling strengths. For smaller values of $g$, the results agree with the theoretical weak coupling prediction from Eq. \eqref{one_mode_h_ind}. When the induced field grows, the dynamics of $\moy{\sigma^z(t)}$ increasingly deviates from its free solution and therefore also the induced field deviates from this analytical prescription (cf. the orange curves in Fig. \ref{fig:one_mode}). This is even more pronounced for the green curve in Fig. \ref{fig:one_mode} b), where the induced field strongly opposes the external driving field such that it is not justified anymore to compare this to the solution derived under the assumption that the spin almost follows its free evolution. The dynamo effect then breaks down and the induced field does not increase further. This breakdown occurs when the induced field is of the same order of magnitude as the external field and opposing it. Increasing $g$ even further (cf. the red curve in Fig. \ref{fig:one_mode}), initially the induced field is larger than the external field and the spin dynamics are thus almost frozen in this case. The dynamo effect is then not even initiated.

\begin{figure*}
    \includegraphics[width=\linewidth]{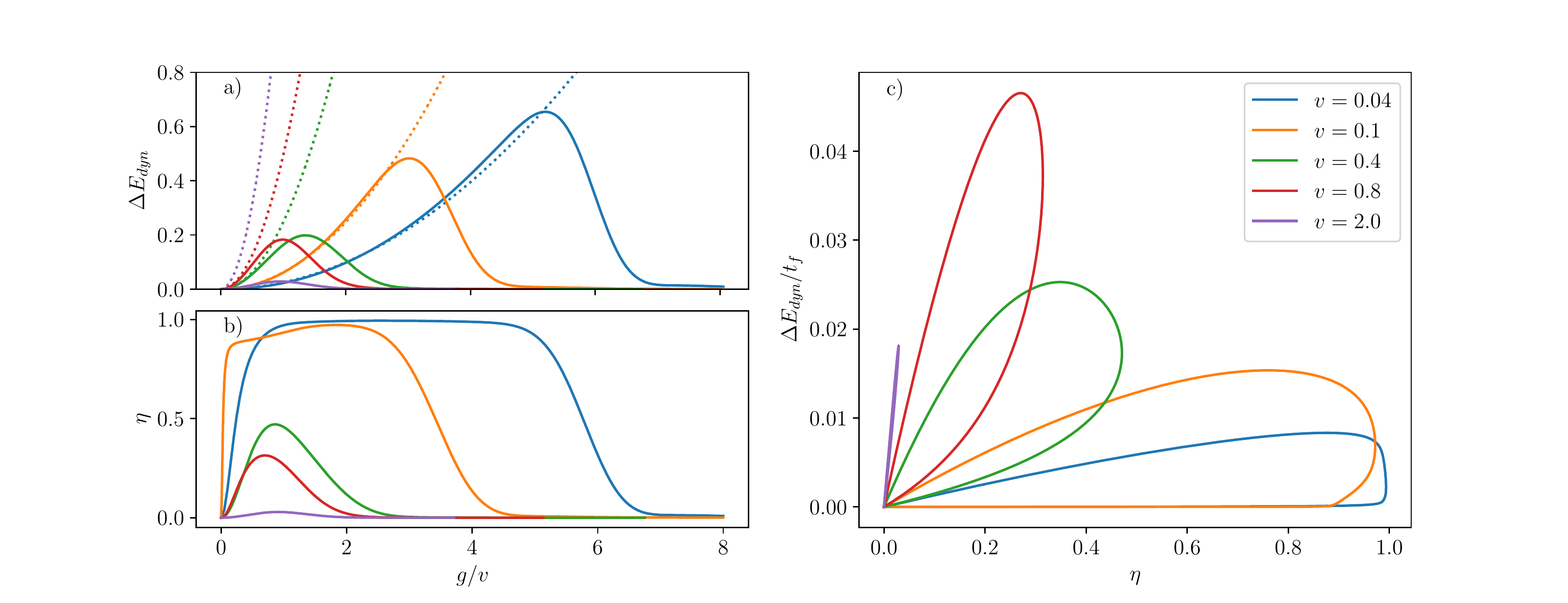}
    \caption{Performances of the one mode dynamo. a) Plot of $\Delta {E}_\text{dyn}$ for different velocities as a function of $g/v$ at $t_f=\pi/v$.
    The solid lines show results from ED, while the dotted line shows Eq. \eqref{eq:E_dyn_half_analytical}. b)
    The corresponding efficiencies $\eta$ as a function of $g/v$.  c) Parametric plot of $\eta$ vs $\Delta {E}_\text{dyn}/t_f$ (averaged output power) for different velocities and with a parameter $g/v$ at $t_f=\pi/v$.
    The ED results are based on a simulation of the time evolution under Hamiltonian Eq. \eqref{eq:single_mode_hamiltonian} with $H = 1.0$ and starting from preparation (1). The Hilbert space of the bosonic mode is truncated for each data point at a value well beyond the expectation value of the filling.
    }
    \label{fig:Edyn_eff_one_mode}
\end{figure*}

From the ED results, we can also numerically evaluate observables of the bosonic mode and thereby classify the performance of the dynamo.
In Fig. \ref{fig:Edyn_eff_one_mode} a) we show the dynamo energy $\Delta E_\text{dyn}$ after half a period of driving (i.e. at $t=\pi/v$) as a function of $g/v$ for different driving velocities.
From here it is clear that Eq. \eqref{eq:one_mode_E_dyn} derived for weak coupling and adiabatic driving quantifies the dynamo energy correctly until the breakdown of the dynamo.
In Fig. \ref{fig:Edyn_eff_one_mode} b), we show the corresponding efficiencies. In Fig. \ref{fig:Edyn_eff_one_mode} c), we plot the averaged output power $E_\text{dyn}/t_f$ parametrically as a function of the efficiency, where the parameter is $g/v$.
We see that the dynamo can reach large efficiencies. Note that the definition from Eq. \eqref{eq:one_mode_E_dyn} here holds for arbitrary times (and not only for full periods), as the ground state energy of the spin without coupling to the bath is invariant in time. There is a well-defined maximum output power $\Delta {E}_\text{dyn}/t_f$ at a non-optimal efficiency. Beyond that, corresponding to an induced field growing too large, the spin becomes polarized due to the large field coming from the shifted modes and both the averaged output power and the efficiency go to zero. The maximum output power first increases with the velocity, until for large velocities the dynamo breaks down and both the efficiency and the output power diminish. \par
We note that for a fixed $g$, decreasing $v$ too much also results in a breakdown of the dynamo effect: On the one hand, since $h(0) = -2g^2/v$ in preparation (1), it can result in a large initial field which leads to the static spin limit. On the other hand, decreasing $v$ also increases the time of the experiment (e.g. in Figs. \ref{fig:Edyn_eff_one_mode} we are considering driving the system from $t=0$ to $t=\pi/v$). This then leads to a smaller output power averaged over the time of the drive. This is illustrated in Fig. \ref{fig:Edyn_eff_one_mode} c).
\par
Note finally that the analytical Eqs. \eqref{eq:one_mode_work} and \eqref{eq:one_mode_E_fluct} hold only in the limit of unitary and adiabatic spin dynamics and are thus not suited to describe the plot in Fig. \ref{fig:Edyn_eff_one_mode} c). Indeed for $t=\pi/v$, we would find $\eta = 1$ independently of the other system parameters. Therefore, these equations describe only a special regime, which in Fig. \ref{fig:Edyn_eff_one_mode} c) corresponds to the blue curve at intermediate parameter $g/v$ (i.e. the region where the efficiency is close to one for a range of $\Delta {E}_\text{dyn}$).
A numerical analysis of the energy of fluctuations is presented in appendix~\ref{app:fluctuations_1ED}.

In the following, we want to examine the situation when we couple to a bigger number of bosonic modes whose frequencies are given by a broad spectrum.

\subsection{Finite number of bath modes}\label{sec:finite_number_modes}
To depart from coupling to one single bosonic mode only, let us briefly discuss an intermediate step towards coupling to a continuous bath, which is a finite number of discrete bosonic modes. Considering a dozen of modes, one can still use numerical simulation based on ED to get full information on the bath evolution, while observing the dynamo from the spin observables.
To characterize the discrete bath and compare with the continuous case analyzed below, we fix the couplings $g_k$ from a linear (Ohmic) spectral density with a hard high-frequency cut-off $\omega_\text{c}$:
\begin{equation}\label{eq:hard_cutoff_spectral_density}
    J(\omega) = \begin{cases}
2 \pi \alpha \omega & \omega \leq \omega_\text{c} \\
0 & \omega > \omega_\text{c}.
\end{cases}
\end{equation}
We are considering the so called scaling limit where $\omega_\text{c}$ is the largest energy scale of the problem and in particular $\omega_\text{c} \gg H$ \cite{orth2013nonperturbative}.
Discretizing the spectral density, we identify $g_k^2 = 2 \alpha \omega_k \Delta \omega_k$ where $\Delta \omega_k$ is the width of the part of the spectrum that is represented by the mode $k$.

\begin{figure*}
    \centering
    \includegraphics[width = \linewidth]{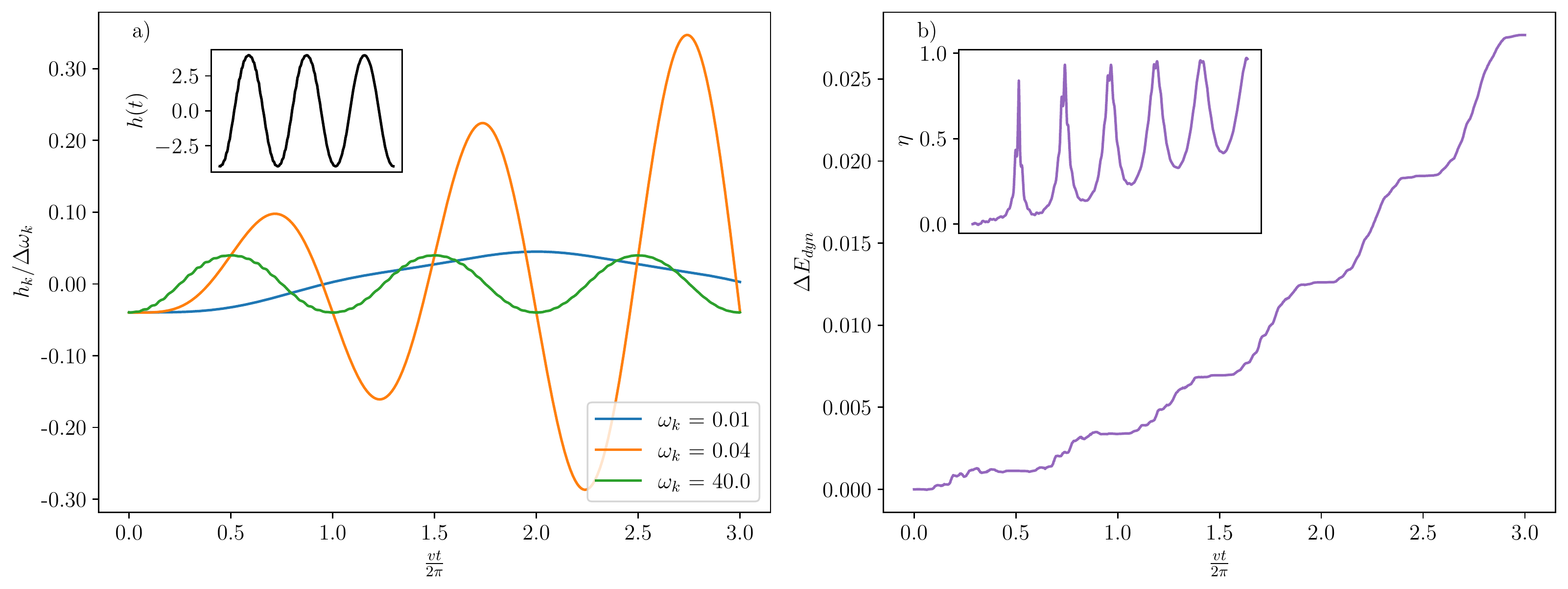}
    \caption{Results from ED for a spin coupled to twelve modes with frequencies from 0.0 to 100.0 and a (discretized) linear spectral function with cut-off at $\omega_\text{c} = 100.0$ and a small coupling corresponding to $\alpha = 0.02$. The spin is driven with velocity $v = 0.04$ and the magnitude of the external field is $H=1.0$. The Hilbert space of each mode is truncated above its maximal occupation. a) The induced field for three different modes. In the inset, the summed induced field is shown which is dominated by the high frequency modes. b) The dynamo energy $\Delta E_\text{dyn}$ and in the inset the efficiency $\eta$.}
    \label{fig:manymode_ED}
\end{figure*}

For modes with $\omega_k \gg v$, we still expect that they simply follow the spin due to their fast relaxation, i.e. $h_k(t) = -g_k^2/\omega_k \langle \sigma^z(t) \rangle$ (similar to Eq. \eqref{eq:one_mode_w_gg_v} for adiabatic driving).
At resonance $\omega_k = v$, for small values of the couplings $g_k$, assuming the free solution of $\langle \sigma^z(t) \rangle$ in Eq. \eqref{eq:Usz} we can solve Eq. \eqref{eq:ht1} exactly for any velocity and find a term $\propto g^2 v t \sin(vt)$ as a contribution to the induced field, similarly as in Eq. \eqref{one_mode_h_ind}.
This resonant contribution can also be seen from the numerical results in Fig. \ref{fig:manymode_ED} a).
Since the resonant region with $\omega_k \sim v$ is small and the largest part of the spectrum just tends to follow the spin and induce a field $h_k(t) = -g_k^2/\omega_k \langle \sigma^z(t) \rangle$, we then expect only a small influence of the resonant dynamically induced field on the total induced field. This behaviour is also confirmed from ED with twelve modes and a linear spectrum (see the inset of Fig. \ref{fig:manymode_ED} a): the total induced field roughly follows the spin dynamics). Since the induced field coming from the high-frequency modes does not oppose the external driving field, it can effectively stabilize the dynamo field which would otherwise have a strong influence on the spin dynamics and thereby lead to a breakdown of the dynamo effect.\par
We can apply the definitions of the quantity $\Delta E_{\text{dyn}}$ from Eq. \eqref{eq:E_dyn} and of the work done by driving $W_\text{dr}$ from Eq. \eqref{eq:drive_work_integral} and analyse the performance of the dynamo.
The analysis of $\Delta E_{\text{dyn}}$ and the efficiency $\eta$ is shown in Fig. \ref{fig:manymode_ED} b) for a fixed coupling strength.
We observe after every half period (i.e. at times $t=n \pi/v$ with $n \in \mathbb{N}$), the dynamo still reaches large efficiencies (inset of Fig. \ref{fig:manymode_ED} b)) and the depth of the intervening valleys is decreasing for longer times.
This result supports the claim of stabilization of the dynamo effect for coupling to a broad spectrum: Driving the system for several periods, instead of breaking down (as it occured for coupling to a single resonant mode), the efficiency now approaches unity. This is consistent with the fact that the dynamo energy in this regime increases and dominates the energy of fluctuations, which we analyze in Appendix~\ref{app:fluctuations_NED}.

The results for a discrete spectrum hint how the dynamo effect can be understood for an Ohmic spin boson model (in which the bath consists of infinitely many modes and the spectral function becomes continuous): The near resonant modes still induce a dynamical field which opposes the change of the external field. However, as opposed to the effect we observed for one mode, the dynamically induced field now does not immediately obstruct the spin dynamics, which is protected by the dominant contribution from other bath modes and thus follows its unitary dynamics for much longer. In the following, we will make this more explicit for the case of a continuous bath.

\section{The quantum dynamo for coupling to a continuous bath}\label{sec:multimode}

In this section we are investigating how the quantum dynamo effect may occur also for a spin coupled to a large number of bosonic modes according to the Hamiltonian in Eq. \eqref{d:model}. This corresponds to a periodically driven spin-boson model \cite{grifoni1995cooperative, grifoni1996exact, grifoni1998driven}. In this case, the induced field can be computed for each mode using Eq.~\eqref{eq:one_mode_ODE}, with the important difference that the dynamics of $\moy{\sigma^z(t)}$ is now influenced by the ensemble of modes. To set the stage, we recall some general results for the Ohmic spin-boson model and extend them to the special case we are considering in this paper.

\subsection{General results for Ohmic spin-boson model}\label{sec:multimode_general}
Let us first summarize a few general results for the spin-boson model from the well established literature.
\par
The evolution of the spin under influence of the bath can be expressed exactly using the influence functional method developed by Feynman and Vernon \cite{feynman1963theory} as a double sum over all spin paths. The influence of the bath is in this formulation given by the so called influence functional \cite{feynman1963theory, leggett1987dynamics, orth2013nonperturbative}:
\begin{equation}\label{eq:spin_path_integral}
  \braa{\sigma_f} \rho_s(t) \kett{\sigma'_f} = \int \mathcal{D}\sigma \int \mathcal{D}\sigma' \mathcal{A}[\sigma]\mathcal{A^*}[\sigma'] \mathcal{F}[\sigma,\sigma'].
\end{equation}
On the left hand-side, we have a component of the spin density matrix $\rho_s(t)$ where $\kett{\sigma_f}, \kett{\sigma_f'} \in \{\kett{\uparrow}, \kett{\downarrow} \}$. On the right hand side, we evaluate this element as a double sum over all spin parts in the configuration space of the spin in $z$-basis. This accounts for flips of the spin, so that here $\sigma, \sigma' = \pm 1$. The influence of the free spin part $\mathcal{H}_\text{spin}$ defined in Eq.~\eqref{d:system} is encapsulated in the amplitudes $\mathcal{A}[\sigma]$ through Eq.~\eqref{eq:DsigmaAsigma}. The influence of the harmonic bath is then described by the influence functional $\mathcal{F}[\sigma,\sigma']$ defined in Eq~\eqref{eq:influence_functional}.
More details about this formulation in relation with a numerical method that is based upon it and of which we will present results in the following can be found in Appendix \ref{app:path_integral}. \par
The path integration in Eq. \eqref{eq:spin_path_integral} cannot be performed analytically and so one has to either try and tackle it numerically through a stochastic approach (see Appendix \ref{app:SSE}) or use approximations. A famous approximation to derive simpler expressions for the spin evolution is the so called non-interacting blip approximation (NIBA) \cite{leggett1987dynamics}.\par
It is based on the decomposition of spin paths into so called blips and sojourns: In the double path integral description of the spin dynamics, a sojourn corresponds to a diagonal state, while a blip corresponds to an off-diagonal state. Through the influence functional, a blip is coupled to all previous blips and sojourns, leading to a non-Markovian problem \cite{leggett1987dynamics, orth2013nonperturbative}. The NIBA then corresponds to neglecting the interblip correlations and sojourn-blip interactions beyond neighboring ones \cite{leggett1987dynamics,grifoni1998driven}. This is justified when the average blip length is much shorter than the average sojourn length \cite{leggett1987dynamics,orth2013nonperturbative}, which can be verified self-consistently for a given problem \cite{grifoni1998driven}. The parameter ranges in which the NIBA gives reliable results are quite restricted and disconnected and depend on the nature of the problem.
\par At zero temperature and in the unbiased case (i.e. no field in $z$-direction acting on the spin) with only a constant field in $x$-direction and for $0 < \alpha \leq \frac{1}{2}$, the NIBA interestingly correctly predicts a renormalization of the tunneling element as
\begin{equation}\label{eq:renormalized_tunneling}
 \Delta_r = \Delta \left(\frac{\Delta}{\omega_\text{c}} \right)^\frac{\alpha}{1-\alpha}.
\end{equation}
Furthermore, in this limit the NIBA correctly predicts the quality factor of the damped oscillation in $\moy{\sigma^z(t)}$ for short to intermediate times, but fails to correctly predict the long-time behaviour \cite{leggett1987dynamics,grifoni1998driven,orth2013nonperturbative}. \par
In general, the crucial criterion for the NIBA to work is of course that the average blip length is short. This applies at high temperatures, strong damping or a large bias field \cite{leggett1987dynamics, grifoni1995cooperative, grifoni1997coherences, orth2013nonperturbative}. Therefore, we can assume for our periodically driven setup that the NIBA will apply at short times when the external field in $z$-direction is large, while it is expected to break down when approaching $t=\pi/(2v)$ where the external field is pointing in $x$-direction. However, inducing a large field in $z$-direction can justify the NIBA in a wider range and we will see below how this can occur in our setup.\par
The NIBA approximation can be applied directly to the spin dynamics given by an influence functional in the path integral formulation \cite{leggett1987dynamics}. There is an intuitive way to understand the NIBA in this case from a polaronic transformation and subsequent replacement with bath operators by their free evolutions \cite{dekker1987noninteracting}. A similar method has also been applied to derive the spin dynamics under the NIBA with a time-dependent bias field \cite{dakhnovskii1994dynamics}. A generalization for a setup where both the bias and the tunneling element acting on the spin are time-dependent can also be derived \cite{grifoni1996exact, grifoni1997coherences, grifoni1998driven} and is given in the Appendix \ref{app:NIBA}.
\par
Beyond the NIBA, one can employ a mapping to the anisotropic Kondo model \cite{guinea1985bosonization} which has been solved by the Bethe ansatz \cite{ponomarenko1993resonant}. This allows to make predictions about the spin expectation values in the spin-bath ground state in different limits of the spin-boson model. While the general solutions from the Bethe ansatz take quite complicated forms, they admit simplifications in certain limits that have been derived in \cite{kopp2007universal}. In particular, when $1-\alpha \gg \Delta/\omega_\text{c}$ where $\Delta$ is the tunneling element (i.e. $\Delta = H \sin(vt)$) and if the driving is slow so that we can effectively set $vt = \theta = cst.$ and therefore $\Delta$ is fixed, from correspondence to the anisotropic Kondo model it can be shown that the low energy physics is controlled by the Kondo scale $T_K = \Delta \left(\Delta/D \right)^{\alpha/(1-\alpha)}$ \cite{kopp2007universal} which in its form reminds the renormalized tunneling element that is also found in the NIBA. The difference is in the high-energy cut-off $D$ which is related to $\omega_\text{c}$ by \cite{cedraschi2000zero}
\begin{equation}
    \left( \frac{D}{\omega_\text{c}}\right)^{2\alpha} =  \frac{2 \Gamma(3/2 - \alpha) e^{-b}}{\sqrt{\pi} (1-2\alpha) \Gamma(1-2\alpha) \Gamma (1-\alpha)},
\end{equation}
with $b=\alpha \log(\alpha) + (1-\alpha) \log(1-\alpha)$.\par
\begin{figure}
    \centering
    \includegraphics[width = 0.45\textwidth]{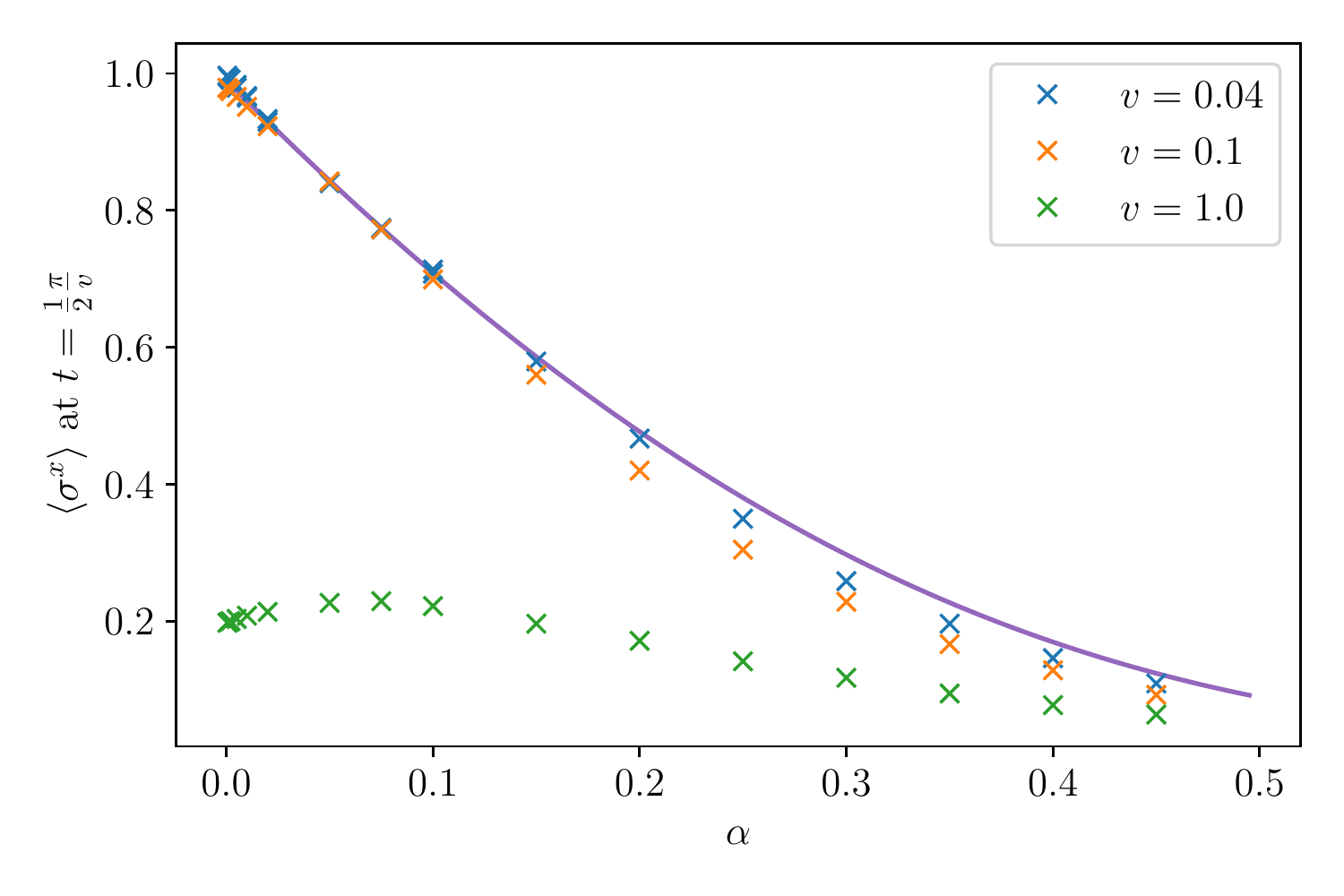}
    \caption{Comparison of SSE result at $t=\pi/(2v)$ (crosses, different colors represent different velocities) with result from Bethe ansatz Eq. \eqref{eq:sx_bethe} (purple line) for different velocities. We see that for quasi-adiabatic velocities, the value of $\moy{\sigma^x}$ at $t=\pi/(2v)$ agrees with the equilibrium theory, while for larger velocities and coupling strengths there are non-adiabatic effects. }
    \label{fig:bethe}
\end{figure}

In the limit where the bias is small compared to the Kondo scale, one finds \cite{kopp2007universal}
\begin{equation}\label{eq:sx_bethe}
    \moy{\sigma^x} \approx \frac{1}{2 \alpha -1} \frac{\Delta}{\omega_\text{c}} + C_1(\alpha) \frac{T_K}{\Delta},
\end{equation}
where $C_1(\alpha) = \frac{e^{-b(2-2\alpha)}}{\sqrt{\pi}(1-\alpha)} \frac{\Gamma(1-1/(2-2\alpha))}{\Gamma(1-\alpha/(2-2\alpha))}$. In our model, a small bias is achieved at $vt= \pi/2 + n\pi$ with $n \in \mathbb{N}$. Therefore, with adiabatic driving speeds we can compare Eq. \eqref{eq:sx_bethe} with the spin expectation value
$\moy{\sigma^x(t = \frac{\pi}{2v})}$ and compare it with the numerical stochastic approach in Fig. \ref{fig:bethe}.

\subsection{Weak coupling to the bath: GKLS master equation}\label{sec:GKLS}

To analyze the long time evolution, we focus on the weak-coupling limit, which allows us to use well-known techniques to access the approximate spin dynamics \cite{breuerbook}. We here focus on the GKLS master equation. In addition to providing precious insights about the long time evolution of the quantum dynamo, this approach has the advantage to provide analytical expressions of the physical quantities of interest. The GKLS equation can be derived for the present model when $J(\Omega \pm v),\,J(v) \ll k_B T \ll  v,H,\omega_\text{c}$. Here, $\Omega = \sqrt{H^2+v^2}$ is the effective spin frequency in the frame rotating at the driving frequency, as introduced in Sec.~\ref{sec:case_of_study} (i.e. the energy splitting of Hamiltonian Eq. \eqref{eq:effective_spin_Hamiltonian}). $T$ is the bath temperature which in this section is assumed to be non-zero, but still much smaller than the spin and driving frequencies. This assumption is necessary to ensure a correlation time of the bath much shorter than the dissipation time-scale, as required by the GKLS framework. In the present case, this essentially means that the low-frequency modes are treated differently from the other approaches we used in the article (they are assumed to be highly populated). While this difference of treatment certainly contributes to the discrepancies we observe between the numerical predictions of the GKLS spin dynamics and the other numerical methods we employ, the error appears small at weak enough coupling, see Fig. \ref{fig:numerics_comparison}.

The GKLS equation can be derived applying standard tools of quantum open system theory (See Appendix \ref{app:GKLS}), including the Born-Markov and secular approximations. Moreover, any back-action on the system dynamics is neglected.
As the spin is driven periodically, we obtain what is sometimes called a Floquet-Markov master equation \cite{Grifoni98}, describing the relaxation and dephasing dynamics between pure-state periodic orbits of the spin (which would be stable in the absence of coupling to the reservoir).

Here, the two periodic orbits are described by $\ket{\Psi_-(t)} =$
\bb
\bigg(\sqrt{\frac{\Omega+H}{2\Omega}}\cos\left(\frac{vt}{2}\right)-i\sqrt{\frac{\Omega-H}{2\Omega}}\sin\left(\frac{vt}{2}\right)\bigg)\ket{\uparrow}\nonumber\\
+ \bigg(i\sqrt{\frac{\Omega-H}{2\Omega}}\cos\left(\frac{vt}{2}\right)+\sqrt{\frac{\Omega+H}{2\Omega}}\sin\left(\frac{vt}{2}\right)\bigg)\ket{\uparrow},\nonumber\\\label{eq:orbit}
\ee
and the state $\ket{\Psi_+(t)}$ which is the spin state orthogonal to $\ket{\Psi_-(t)}$.

The GKLS master equation takes the form:
\bb
\dot \rho = -i[{\cal H}_S(t)+{\cal H}_\text{LS}(t),\rho] + {\cal L}_\text{relax}(t)\rho + {\cal L}_\text{deph}(t)\rho,\label{eq:GKLS}\nonumber\\
\ee
where we have defined a Lamb shift correction ${\cal H}_\text{LS}(t)$ to the free spin Hamiltonian induced by the coupling to the bath, as well as two dissipative contributions ${\cal L}_\text{relax}$ and ${\cal L}_\text{deph}$. They read
\begin{widetext}
\bb
{\cal L}_\text{relax}(t) &=& \gamma_\text{relax}{\cal D}[\ket{\Psi_-(t)}\bra{\Psi(t)}],\\
{\cal L}_\text{deph}(t) &=& J(v)\frac{H^2}{4\Omega^2}{\cal D}\bigg[\ket{\Psi_+(t)}\!\bra{\Psi_+(t)}-\ket{\Psi_-(t)}\!\bra{\Psi_-(t)}\bigg],
\ee
\end{widetext}
where we have introduced the dissipation superoperator acting as ${\cal D}[X]\rho = X\rho X^\dagger -\tfrac{1}{2}(X\dagger X\rho-\rho X^\dagger X)$.
The superoperator ${\cal L}_\text{relax}(t)$ captures relaxation to the periodic orbit $\ket{\Psi_-(t)}$ which corresponds to the stationary, long-time spin dynamics. The relaxation mechanism can be attributed to the bath modes at frequencies close to $\Omega\pm v$. Indeed, the relaxation rate $\gamma_\text{relax} = \sum_{l=\pm}(\Omega-lv)^2J(\Omega+lv)/4$ is proportional to the spectral density of the reservoir evaluated at these two frequencies.
Secondly, the Lindbladian ${\cal L}_\text{deph}(t)$ is responsible for pure-dephasing in the basis $\{\ket{\Psi_-(t)},\ket{\Psi_+(t)}\}$ and can be attributed to the interaction with bath modes at frequency $v$. \par
We emphasize that these relaxation mechanisms are combined effects involving both the bath modes and the drive, which allow to emit excitations at frequencies different from the spin transition frequency $H$.
This mechanism is reminiscent of the radiative cascade occurring of a resonantly driven two-level atom coupled to the electromagnetic vacuum \cite{cohentannoudjibook} and is more generally typical for periodically driven quantum open systems. In this context, the periodic drive can be thought of as a bosonic mode containing a large number of excitations able to assist transitions between two joint system-reservoir states by providing a quantum of energy $\hbar\nu$. \\
The dynamics along the periodic orbit can be used to evaluate the induced field at long time and the associated dynamo energy. It is convenient to express this periodic motion in terms of the Bloch coordinates. We have:

\bb
\moy{\sigma^x(t)} &=& \frac{H}{\Omega}\sin(vt),\\
\moy{\sigma^y(t)} &=& \frac{v}{\Omega},\\
\moy{\sigma^z(t)} &=& \frac{H}{\Omega}\cos(vt). \label{eq:stationary_sz}
\ee
Note that the stationary orbit is very close to the free unitary spin evolution (in the absence of a bath) of Eqs. \eqref{eq:Usx}, \eqref{eq:Usy} and \eqref{eq:Usz} as long as $v\ll H$. However, here the dissipation induced by the bath is taken into account, and actually plays a key role in stabilizing this orbit from any initial condition of the spin. For larger velocities, the asymptotic value of $\moy{\sigma^y}$ increases, while the diameter $H/\Omega$ of the periodic circle followed in the Bloch sphere decreases.

\subsection{Numerical results for the spin dynamics}

\begin{figure*}
    \centering
    \includegraphics[width=\textwidth]{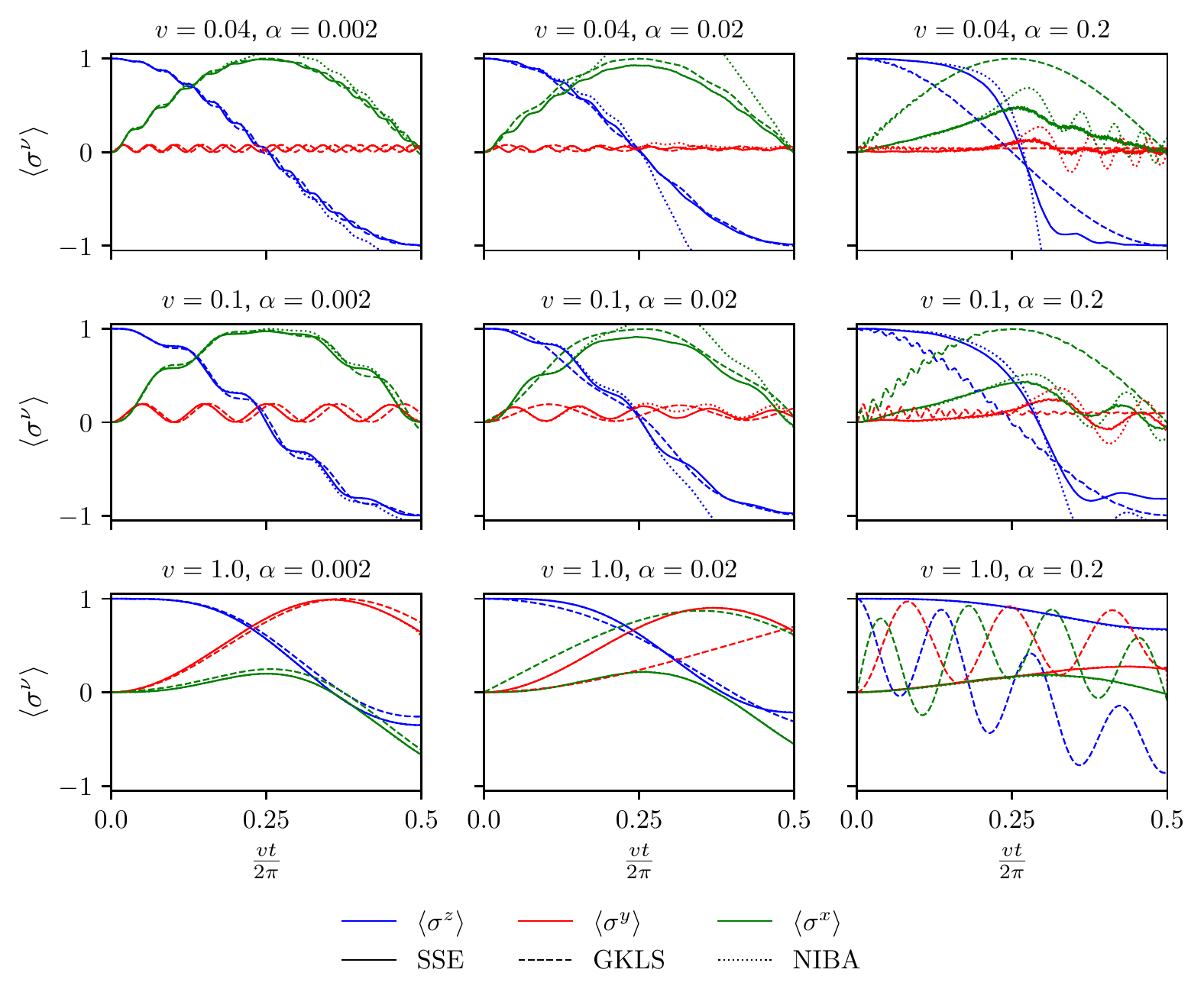}
        \caption{Comparison of numerical results. The expectation values of $\moy{\sigma^z}$, $\moy{\sigma^y}$ and $\moy{\sigma^x}$ are shown in blue, green and red in this order. Solid lines represent results from the stochastic approach, dashed lines represent results from the Floquet approach and dotted lines represent results from the NIBA. Here, $H=1.0$.}

        \label{fig:numerics_comparison}
\end{figure*}
To evaluate the spin dynamics in the driven spin-boson model we consider, we employ the numerically exact stochastic approach developed in \cite{orth2013nonperturbative}, referring to the stochastic Schr\"odinger equation (SSE) approach. For $\alpha<1/2$ it allows us to evaluate the spin dynamics.
A comparison of these results for the spin expectation values to the theoretical prediction from the NIBA and the GKLS master equation approach (cf. Sec. \ref{sec:GKLS}) is shown in Fig.~\ref{fig:numerics_comparison} where results from the stochastic approach are represented by solid lines. At quasi-adiabatic to intermediate driving speeds and at small coupling $\alpha$, the spin follows the drive and shows secondary oscillations with period $\Omega$. The latter are with time increasingly suppressed which leads to a smoothing of the spin dynamics. The expectation value $\moy{\sigma^y}$ approaches a constant value. Qualitatively, this behaviour agrees with the predictions from the Floquet master equation. The results from the NIBA agree for small times, when the external field has not yet been fully oriented in $x$-direction ($t < \pi/(2v)$). At $t = \pi/(2v)$, the external field points in $x$-direction and we find $\moy{\sigma^x}<1$, which among the approaches presented here is predicted analytically only by the mapping with the Kondo model. This gives Eq.~\eqref{eq:sx_bethe}, which explicitly captures the entanglement of the spin-bath ground state. In contrast, such an entanglement is neglected within the GKLS equation which assumes weak system-bath coupling. The comparison of the results from the stochastic approach to this formula is shown in Fig. \ref{fig:bethe} for different velocities, which demonstrates remarkable agreement at quasi-adiabatic driving speed.\par
For higher values of $\alpha$, the dynamics revealed by the stochastic approach differ: Secondary oscillations in $\moy{\sigma^z}$ are smoothed out and the shape of the curve is altered. Intuitively, this can be understood by the changed ground state with the renormalized tunneling element (given in Eq. \eqref{eq:renormalized_tunneling}) due to the bath. Navigating adiabatically from the north to the south pole with a Hamiltonian
\begin{equation*}
\mathcal{H} = -\frac{H}{2}\cos(vt)\sigma^z - \frac{\Delta_r(t)}{2} \sigma^x,
\end{equation*}
with $\Delta_r(t) = H \sin(vt)\left(\frac{H \sin(vt)}{\omega_\text{c}} \right)^\frac{\alpha}{1-\alpha}$,
we can find the expectation value $\moy{\sigma^z}$ from the adiabatic ground state which leads to a qualitatively correct prediction in the range of $\alpha$ at question. Note that from the Heisenberg equation of motion, we have $\moy{\dot{\sigma}^z(t)} = -H\sin(vt)\moy{\sigma^y(t)}$. It is interesting to observe that $\moy{\sigma^y}$ stays close to zero initially due to a smoothing of secondary oscillations in $\moy{\sigma^z}$ corresponding to a suppression of non-adiabatic effects, but then exhibits a ``bump'' around time $t = \pi/2v$.
The expectation value $\moy{\sigma^x}$ approaches its unbiased value described by Eq.~\eqref{eq:sx_bethe} in an almost linear fashion which close to the north pole is captured by the NIBA solution.

\par
Finally, for higher velocities $v=H$, at small values of $\alpha$, the dynamics of $\moy{\sigma^z}$ is captured by the Floquet approach and agrees with the results of the stochastic approach. At higher values of $\alpha$ the two approaches start to differ but interestingly the NIBA now makes matching predictions throughout the half period of the drive shown in the figure.
One of the situations in which the NIBA is working well is that of a large field in $z$-direction. We can thus hypothesize, that a large field is induced back onto the spin for the entire time of the drive due to the non-adiabatic response at larger velocities. In the following, we will make the nature of the induced field more explicit.

\subsection{Induced field: analytical formula}

Here we study the induced field of the bath modes onto the spin in a similar way as in Sec. \ref{sec:one_mode} for only one mode.
As before, we focus on two different choices of preparations that we generalize to the case of many modes:

\begin{itemize}
    \item Preparation (1): The spin is pointing upward and the bath is in equilibrium with the spin, such that each of the bath modes is in its shifted oscillator ground state at $t=0$, i.e. $h_k(0) = -g_k^2/\omega_k$, $\dot{h}_k(0)=0$.
    \item Preparation (2): The spin is initially pointing upwards. The bath is initially in vacuum and is brought into contact with the spin at $t=0$, i.e. $h_k(0) =0$, $\dot{h}_k(0)=0$.
\end{itemize}
Solving formally Eq.~\eqref{eq:one_mode_ODE} (with $\omega \to \omega_k$), we obtain:
\bb
    h_k(t) &=& -\delta_{(1)} \frac{g_k^2}{\omega_k} \cos(\omega_k t)\nonumber \\&-& g_k^2 \int_0^t dt' \sin(\omega_k(t-t')) \langle \sigma^z(t') \rangle .\label{eq:formal_solution_hk} \\
    &=& \frac{g_k^2}{\omega_k}\bigg((1-\delta_{(1)}) \cos(\omega_k t) - \moy{\sigma^z(t)}\nonumber\\
    &&\quad\quad\quad+ \int_0^t dt' \cos(\omega_k(t-t')) \moy{\dot{\sigma}^z(t')} \bigg) \label{eq:formal_solution_hk_2}
\ee
To go to the second line, we integrated by parts. This allows to identify respectively a free contribution (proportional to $\cos(\omega_k t)$), an adiabatic contribution (proportional to $\moy{\sigma^z(t)}$) and a dynamo contribution coming from the interplay between the non-adiabatic spin dynamics and the mode dynamics. The discussion of the previous section allows us to conclude that the modes resonant with the typical frequency scales of the evolution of $\moy{\dot{\sigma}^z(t')}$ will contribute mostly to the dynamo effect. As before, it is legitimate in the case of very weak coupling to the bath and slow driving to expect an evolution following the ground state, i.e. $\moy{\sigma^z(t)} \sim \cos(vt)$, such that the modes of frequency $\omega_k\sim v$ will convey an efficient dynamo effect.
On the other hand, the spin dynamics and the average performance of the dynamo are related to the total induced field $h(t)$ which is obtained by summing over the modes, and therefore depends on the spectral density $J(\omega)$ of the bath:
\bb\label{eq:formal_solution_ht}
    h(t) &=& h_\text{free}(t) + h_\text{ad}(t) + h_\text{dyn}(t),\\
    h_\text{free}(t) &=& (1-\delta_{(1)})\int_0^\infty d\omega \frac{J(\omega)}{\pi\omega}\cos(\omega t),\\
    h_\text{ad}(t) &=& -\int_0^\infty d\omega \frac{J(\omega)}{\pi\omega}\moy{\sigma^z(t)},\\
    h_\text{dyn}(t) &=& \int_0^t \! dt'\int_0^\infty \! d\omega \frac{J(\omega)}{\pi\omega}\cos(\omega (t-t'))\moy{\dot\sigma^z(t')}.
\ee
As pointed out in the previous section, coupling to several modes allows to stabilize the dynamo. In this section, we will show the emergence of this stability for the case of a continuous bath, pinpointing effects associated with dissipation and strong coupling.

Injecting the Ohmic spectral density into the free contribution to the induced field gives
\bb
 h_\text{free}(t) &=& (1-\delta_{(1)}) \frac{2 \alpha \omega_\text{c}}{1 + (\omega_\text{c} t)^2}.
\ee
Note that $h_\text{free}(t)\to 0$ for $t\gg \omega_\text{c}^{-1}$ so that the free contribution can be neglected after a short time independently of the preparation. We remark that this is the reason we did not consider this contribution in Sec. \ref{sec:general}. Physically, it arises in preparation (2), since the system is not initialized in the ground state and corresponds to a fast equilibration.\par
For the adiabatic contribution, we obtain
\bb
\label{eq:h_from_zt_ad}
 h_\text{ad}(t) &=& -2 \alpha \omega_\text{c} \moy{\sigma^z(t)},
\ee
corresponding to a field following the spin in agreement with Eq.~\eqref{eq:general_ad_h}. Finally, for the dynamically induced field, we obtain
\bb
 h_\text{dyn}(t) &=& \int_0^t \! dt' \frac{2 \alpha \omega_\text{c}}{1 + \omega_\text{c}^2 (t-t')^2} \moy{\dot\sigma^z(t')}.\nonumber
\ee
Note that $(1+\omega_\text{c}^2 t^2)^{-1}$ vanishes for $\vert t\vert$ greater than a few $\omega_\text{c}^{-1}$, and the typical spin evolution occurs on much longer time-scales so that one can finally approximate $\moy{\sigma^z(t')} \simeq \moy{\sigma^z(t)}+(t'-t)\moy{\dot\sigma^z(t)}$ and perform the integration over $t'$. This yields:
\bb
 h_\text{dyn}(t) &=& 2 \alpha \arctan(\omega_\text{c} t) \moy{\dot{\sigma}^z(t)}  -\frac{\alpha}{\omega_\text{c}} \log(1 + \omega_\text{c}^2 t^2) \moy{\ddot\sigma^z(t)}, \nonumber\\
 &\approx& \alpha \pi \moy{\dot{\sigma}^z(t)}. \label{eq:h_from_zt_dyn}
\ee
To go to the last line, we used that $\arctan(\omega_\text{c} t) \sim \pi/2$ for $t\gg \omega_\text{c}^{-1}$ and we neglected the second order term since we assume that $\omega_\text{c}$ is the largest energy scale of the problem.

We can use this formula to evaluate the dynamically induced field directly from the expectation value $\moy{\dot{\sigma}^z}$. In this way, Eq \eqref{eq:h_from_zt_ad} and \eqref{eq:h_from_zt_dyn} provide a simple way to determine the induced field and its adiabatic and dynamically induced part from the spin dynamics.

\subsection{Performances of the dynamo}\label{sec:thermodynamics}

\begin{figure}
    \centering
    \includegraphics[width = 0.48\textwidth]{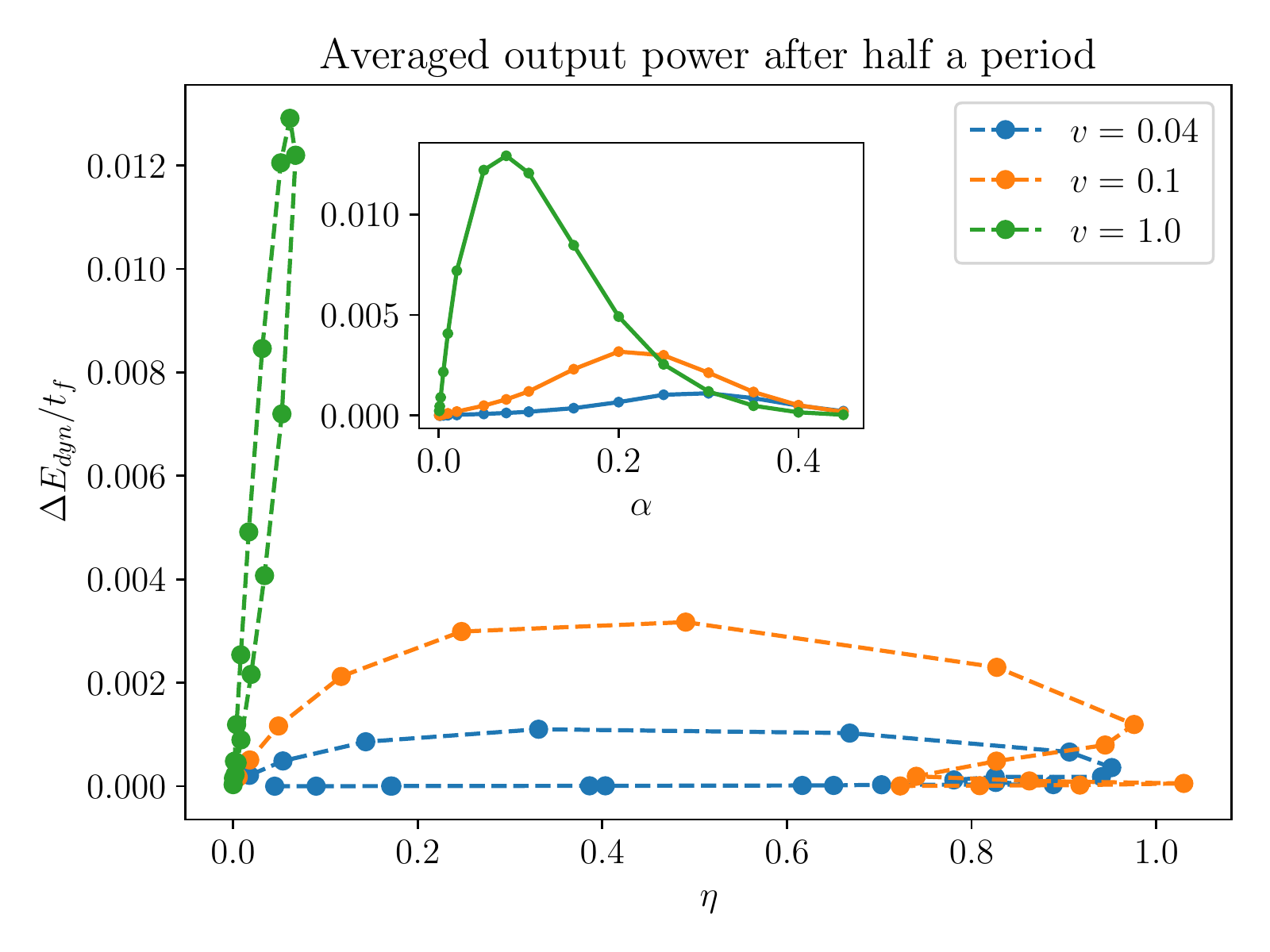}
    \caption{Parametric plot of the averaged output power $\Delta E_\text{dyn}(t_f)/t_f$ at $t_f=\pi/v$ as a function of $\eta = \Delta E_\text{dyn}(t_f)/W_\text{dr}(t_f)$ from results of the stochastic approach. The inset shows the averaged output power as a function of $\alpha$ for comparison. Here $H=1.0$ and $\alpha$ varies between $0.0$ and $0.45$. }
    \label{fig:parametric_SSE}
\end{figure}

Building on the results presented for this setup so far we use the tools developed in Sec. \ref{sec:energetics} to classify the dynamo for a coupling to a continuous bath.
To quantify the dynamo energy, we can use Eq. \eqref{d:dEdyn} with Eq. \eqref{eq:h_from_zt_dyn} and find
\bb
\label{eq:Edyn_continuous}
 \Delta E_\text{dyn} &=& \frac{\alpha \pi}{2} \int_0^t dt' \moy{\dot{\sigma}^z(t')}^2.
\ee
Similarly, the definition of the work done by driving the system in Eq. \eqref{eq:drive_work_integral} is still valid.
This then allows to classify the dynamo effect in terms of the energy transferred into coherent displacement energy due to the dynamical effect and the ratio of the work with this quantity $\eta = \Delta E_\text{dyn}/W_\text{dr}$. Since all of these quantities can be evaluated from the spin expectation values, we can employ the numerically exact stochastic approach to evaluate these quantities. In Fig. \ref{fig:parametric_SSE}, we show results for a similar process as in Fig. \ref{fig:Edyn_eff_one_mode} c) for one mode but now with a continuous bath: Driving the system for half a period (from $t=0$ to $t=\pi/v$), we numerically evaluate the work done to perform the drive and the averaged output power $\Delta E_\text{dyn}(t_f)/t_f$. We then plot the ratio $\eta$ against the averaged output power for different driving velocities. The result is qualitatively similar to the situation of one mode in the sense that for slow driving, we can reach a large conversion ratio at relatively small averaged output power. Increasing the velocity, we can increase the averaged output power at the cost of efficiency of this conversion mechanism. However, comparing Figs. \ref{fig:Edyn_eff_one_mode} c) and \ref{fig:parametric_SSE}, we note that for the same velocity, the maximum output power is higher for the one-mode dynamo than for the Ohmic bath. This is related to a dilution of the effect: When coupling to many modes, the off-resonant modes can consume a fraction of the work done by driving. This is also related to how the effect breaks down when increasing the coupling strength.
Recall that for one mode, the dynamically induced field by a resonant mode becoming too strong fixes the spin, consequently leading to a breakdown of the dynamo effect.
In contrast, for coupling to a continuous bath, we can see from Eq.~\eqref{eq:h_from_zt_dyn} that the dynamically induced field is proportional to $\alpha$ and will thus remain small when the spin dynamics is smooth. Instead, in this case the spin gets fixed by creating strong entanglement with the bath, as pointed out in \cite{henriet2017topology}. In Appendix~\ref{app:fluctuations_SSE} we show from numerical results that for stronger coupling the fluctuation energy can grow larger than the dynamo energy and contribute to the breakdown of the effect. Both for coupling to one resonant mode, as well as to a bath, the effect breaks down when the spin expectation value is fixed for a full period of the driving. As we will describe in the next section, this bridges with dynamically measured topological properties of the spin-$\frac{1}{2}$.
The different mechanisms of breakdown of the dynamo effect as compared to the one-mode situation highlight an operational advantage of the continuous bath dynamo: When operating the dynamo for a longer time, the spin dynamics is protected by the field of the high-frequency modes which tend to follow the spin and by that keep the influence of the dynamically induced field small. While for coupling to one mode, a large dynamo field could build up over time (with an amplitude increasing linearly in time), at weak coupling the dynamo with a continuous bath remains operational with a growing field only at the resonant mode.
\par
Since the numerical stochastic approach is based on the sampling of a stochastic field whose properties depend on the bath correlation (through the function $Q_2(t)$, see Appendix \ref{app:SSE} for details) and consequently requires to solve a stochastic Schr\"odinger equation in time, it is computationally expensive for long times. Therefore, for the long time limit we resort to the Floquet master equation whose derivation we outlined in Sec. \ref{sec:GKLS}.
It is well suited to describe the long time energy balance.
We first compute the work performed by the drive. As explained above, the mechanism responsible for energy exchange with the drive on the long timescale captured by the Floquet master equation is the exchange of quanta of energy $\hbar v$ associated with each emission an excitation into the bath. When the system is in the stationary periodic orbit, the only surviving contribution is associated with the transitions induced by the term ${\cal L}_\text{deph}$. While a jump $\ket{-}\to \ket{-}$ induced by ${\cal L}_\text{deph}$ has no effect on the spin state (it would only induce pure dephasing if the spin had coherences in the $\{\ket{+},\ket{-}\}$ basis), it is accompanied with the scattering of one quantum of energy $\hbar\nu$ from the driving field into the bath. As the rate of these events is $J(v)H^2/8\Omega^2$ in the periodic orbit, one obtains a stationary work flow
\bb
\dot W_\infty = v \frac{H^2}{8\Omega^2}J(v),
\ee
which is equal in magnitude to the energy flow received by the reservoir. One can now seek to determine which fraction of this energy fuels the dynamo effect.

To do so, we inject the periodic orbit motion into Eq.~\eqref{eq:Edyn_continuous} to compute the dynamo energy accumulated during one period:
\bb \label{eq:periodic_Edyn}
 \Delta E_\text{dyn}(t=\pi/v) &=& \frac{\alpha \pi^2 v}{4} \frac{H^2}{\Omega^2}.
\ee
We therefore see that  the long time average efficiency in the weak-coupling limit is unity for any rotation velocity. As a consequence, all the excitations transferred into the bath contribute to displace its modes. We emphasize that this result was obtained under the same approximation we used to derive Eq.~\eqref{eq:h_from_zt_dyn}, assuming that the relevant timescale for the evolution of the spin is $t\gg \omega_\text{c}^{-1}$.\par
In the following section, we will demonstrate, how the properties of the dynamo in the different setups we considered can be linked to the dynamically measured topology of the spin-$\frac{1}{2}$.

\section{Topological properties of the dynamo effect}\label{sec:topology}

\subsection{Topological properties of the spin-\texorpdfstring{$\frac{1}{2}$}{1/2}}

In this Section, we will motivate the particular choice of $\cal H_{\text{spin}}$ in this article to demonstrate the dynamo effect.
Note that the spin part of the model given in Eq.~\eqref{d:system} corresponds to a spin-$\frac{1}{2}$ in a radial magnetic field with a constant offset, i.e.
\bb
    \label{eq:spin_radial_field}
    \mathcal{H}_\text{rad}(\theta,\phi) &=& -\frac{H}{2}(\sin \theta \cos\phi \sigma^x+ \sin \theta \sin \phi \sigma^y + \cos \theta \sigma^z)
    \nonumber\\ &-& \frac{M}{2} \sigma^z,
\ee
where throughout the article we considered $\theta = vt$ and $\phi = 0$, which gives the form in Eq. \eqref{d:system}. This corresponds to a sweep in time of the field along an orbit. Here we introduced for generality the constant offset $M$ in $z$-direction, which throughout the article we set to $M=0$. Its role was in some sense played by the induced field in certain instances. Note that due to the rotational symmetry around the $z$-axis of the Hamiltonian, we can choose $\phi = 0$ without loss of generality and expect similar results for other values of $\phi$. \par
The system described in Eq. \eqref{eq:spin_radial_field} is an example for a topological system \cite{henriet2017topology} and provides an interesting platform for the study of new effects. Recently it has for example been demonstrated that two coupled systems of this form can realize fractional partial Chern numbers \cite{hutchinson2021quantum}.\par
In the realm of the model defined in Eq. \eqref{eq:spin_radial_field} (i.e. neglecting effects from driving and coupling to an environment), the Chern number associated with a state $\ket{\psi}$ can be defined from the Berry curvature \cite{henriet2017topology, hutchinson2021quantum}
\begin{equation}\label{eq:Chern_Berry_curvature}
    C = \frac{1}{2\pi} \int_0^{2\pi} d\phi \int_0^\pi d\theta \mathcal{F}_{\phi \theta},
\end{equation}
where the Berry curvature $\mathcal{F}_{\phi \theta} = \partial_\phi \mathcal{A}_\theta - \partial_\theta \mathcal{A}_\phi$ is defined from the Berry connections $\mathcal{A}_\alpha = i\bra{\psi} \partial_\alpha \ket{\psi}$. It can then be shown that with these definitions Eq. \eqref{eq:Chern_Berry_curvature} can be rewritten for a single spin-$\frac{1}{2}$ with Hamiltonian \eqref{eq:spin_radial_field} as
\begin{equation}\label{eq:Chern_spin_expectation}
    C= \frac{1}{2}\left( \langle \sigma^z(\theta = 0 )\rangle - \langle \sigma^z(\theta = \pi )\rangle \right),
\end{equation}
where the expectation values is taken in the state $\ket{\psi}$. We now focus on the Chern number associated with the ground state of the spin. When $H > M > 0$, we see that $C=1$ as $\moy{\sigma^z(\theta)} = \cos(\theta)$. Conversely, when $M > H >0$ the topology changes and we find $C=0$ as the ground state of the spin points upward at both $\theta = 0$ and $\pi$.\par

During a concrete measurement of the Chern number, a dynamical protocol must be employed in order to sweep through the parameter space of the Hamiltonian \cite{roushan2014observation, Boulder}. Such protocol can be modeled by a Hamiltonian of the form of Eq.~\eqref{d:system} using $\theta=vt$. While the previous discussion can be expected to hold in the limit of vanishing speed $v$, non-adiabatic effects are expected at finite speed $v$. In the present model, the fact that $\langle \sigma^y \rangle$ becomes non-zero signals this non-adiabaticity (cf. Eq. \eqref{eq:Usy} for the situation without coupling to a bath). In such context, it is useful to introduce the \textit{dynamically measured} Chern number which can be evaluated from a measurement of the spin expectation values at $t=0$ and $t=\pi/v$, i.e. \cite{henriet2017topology}
\begin{equation}
\label{eq:Cdynspinexpectation}
    C_\text{dyn} = \frac{1}{2}\left( \langle \sigma^z(t = 0)\rangle - \langle \sigma^z(t = \pi/v )\rangle \right).
\end{equation}
Note that it can be also written as
\begin{equation}\label{eq:Cdyn_sy_integral}
    C_\text{dyn} = \frac{1}{2} \int_0^{\pi/v} dt \moy{\dot{\sigma}^z(t)} = -\frac{H}{2} \int_0^{\pi/v} dt \sin(vt)\moy{{\sigma}^y(t)},
\end{equation}
where the last step follows from the Heisenberg equation of motion for $\sigma^z$ with Hamiltonian \eqref{d:model}, i.e. $\dot{\sigma}^z = -H \sin(vt) \sigma^y$.\par
This dynamically measured Chern number should be interpreted as an estimate of the Chern number accessed using a dynamical procedure. If the spin is not coupled to an environment, non-adiabaticity is characterized by the ratio $v/H$ and one has \cite{gritsev2012dynamical, henriet2017topology}
\begin{equation}
    C_\text{dyn} = C + \order{v/H}.
\end{equation}

When departing from quasi-adiabatic driving speeds, the dynamically measured Chern number $C_\text{dyn}$ starts to deviate from $C$ and thus ceases to be a good estimate for the latter. In \cite{henriet2017topology} it was shown that for a spin coupled to a bath according to Eq. \eqref{d:model} and with an Ohmic spectral density, the bath induces a transition to non-adiabatic behaviour already for coupling strengths well below the critical coupling of the Ohmic spin-boson model. \par
It is therefore clear that $C_\text{dyn}$ is not a quantized number, but takes continuous values due to non-adiabatic effects from driving and entanglement between the spin and the bath. In the light of the dynamo effect, $C_\text{dyn}$ is still an interesting quantity describing the geometry of the dynamically accessed manifold of effective states of the spin. An intuition for this can be gained from the consideration of the breakdown of the dynamo described in Sec. \ref{sec:one_mode_polarized}: When the spin dynamics are frozen because of a strong induced field, the dynamo breaks down and the field is also held fixed, while the dynamo energy is smaller than the energy of fluctuations, as detailed in appendix~\ref{app:fluctuations}. From the definition of $C_\text{dyn}$ in Eq. \eqref{eq:Cdynspinexpectation}, we see that this corresponds to $C_\text{dyn} = 0$. In the following, we are going to make the connection between the occurrence of the quantum dynamo effect and the (dynamically accessed) topology in the studied model \eqref{d:model} more explicit, starting from the case of coupling to one mode.

\subsection{One mode dynamo}

Fig.~\ref{fig:one_mode} b) together with the definition of $C_\text{dyn}$ in Eq.~\eqref{eq:Cdynspinexpectation} can give an intuition of the relation between the dynamo effect and the topology of the spin-$\frac{1}{2}$: From the inset, it appears that a dynamical field opposing the external driving is induced as long as $C_\text{dyn} \sim 1$.

When transitioning to $C_\text{dyn} = 0$, the effect breaks down. From Fig. \ref{fig:Edyn_eff_one_mode} a) and b), where the system is driven from $t=0$ to $t=\pi/v$ corresponding exactly to a sweep from north to south pole, this breakdown can be observed in the dynamo energy $\Delta E_\text{dyn}$ and the efficiency $\eta$.

\begin{figure}
    \centering
    \includegraphics[width=0.45\textwidth]{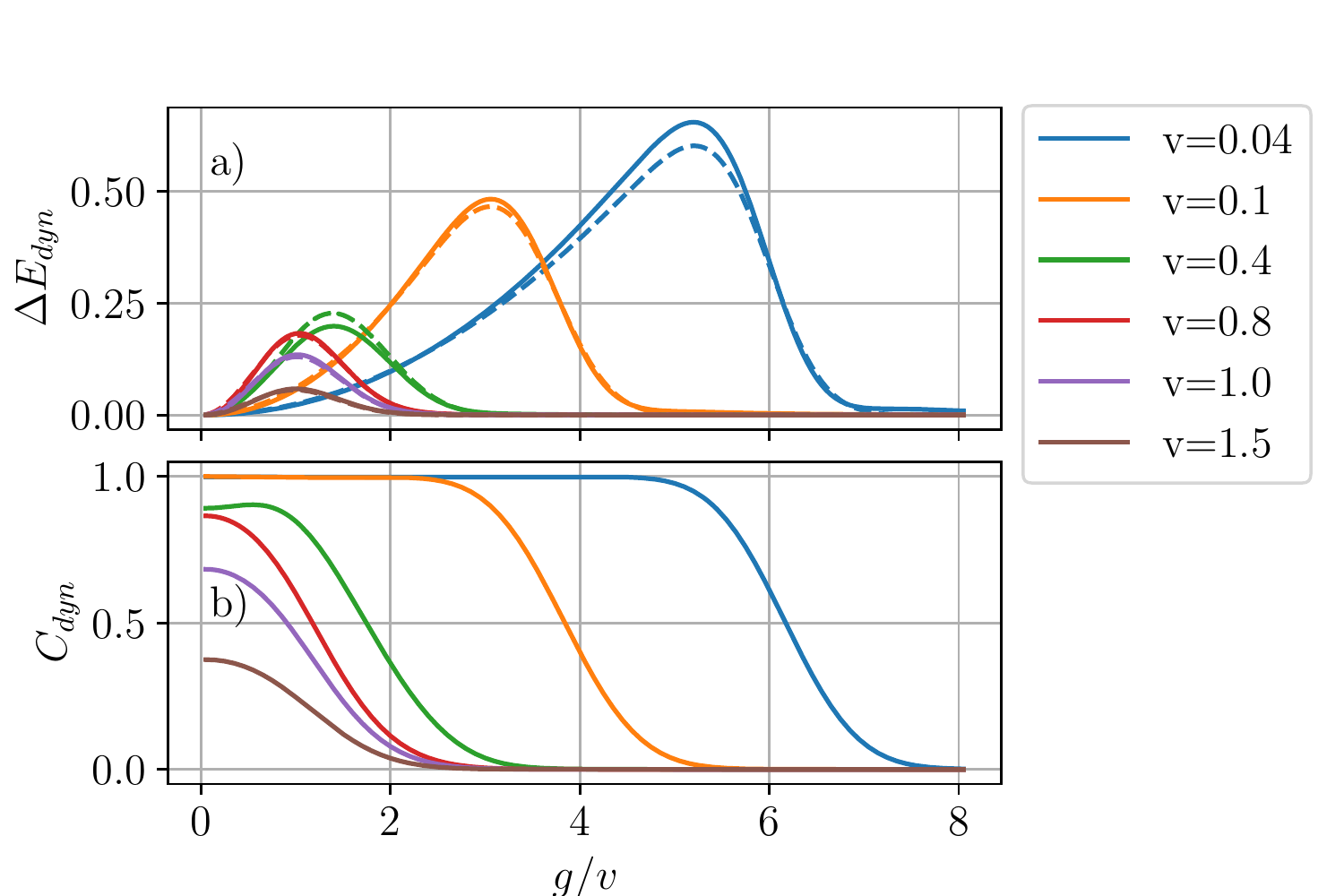}
    \caption{a) The dynamo energy for coupling to a resonant mode from ED simulations (solid lines) and the prediction from Eq.~\eqref{eq:Edyn_Cdyn_one_mode} with $C_\text{dyn}$ measured numerically (dashed lines) as a function of $g/v$. b) The dynamically measured Chern number from the same simulation as a function of $g/v$. For both figures, $H=1.0$, the system was initialized in preparation (1) and the Hilbert space of the bosonic mode is truncated at a value well beyond the expectation value of the filling.}
    \label{fig:Edyn_Cdyn_ED}
\end{figure}

We checked in Fig.~\ref{fig:Edyn_eff_one_mode} a) that at weak coupling, Eq.~\eqref{eq:E_dyn_half_analytical} describes the dynamo energy correctly. Numerically, we find that when driving for one half-period, this equation can be extended to the strong coupling regime as
\begin{equation}\label{eq:Edyn_Cdyn_one_mode}
\Delta E_\text{dyn} =\frac{g^2 \pi^2}{16 v} C_\text{dyn}^2.
\end{equation}
An intuition about the occurence of the factor of $C_\text{dyn}^2$ can be gained from the two limiting cases described in Secs. \ref{sec:one_mode_adiabatic} and \ref{sec:one_mode_polarized}: In the case of weak coupling and adiabatic driving, we have $\moy{\sigma^z(t)} \simeq \cos(v t)$ and $C_\text{dyn} \to C = 1$. The polarized spin limit with adiabatic driving speed corresponds to a situation with a large offset field $M$ (in the language of Hamiltonian~\eqref{eq:spin_radial_field}) and therefore $C_\text{dyn} \to C = 0$ while $\moy{\sigma^z(t)} \simeq 1$. To interpolate between the two cases, we can impose $\moy{\sigma^z(t)} \simeq C(\cos(v t)-1)+1$. This is of course a rough approximation of the adiabatic spin dynamics under the Hamiltonian~\eqref{eq:spin_radial_field} which can be calculated exactly. Nevertheless, since here we are interested in the topological properties which can be encoded from the spin expectation values at the poles which are captured correctly by this form, it can give us some interesting insights. Using Eq.~\eqref{eq:ht1}, we can then approximate the dynamically induced field by
\bb
    h_\text{dyn} \simeq -\frac{1}{2}g^2 vt \sin(v t)C,
\ee
and finally we find
\bb
   \Delta E_\text{dyn}(t=\pi/v) \simeq \frac{g^2 \pi^2}{16 v} C^2.
\ee
Departing from the adiabatic limit, we replace $C$ by $C_\text{dyn}$ and obtain Eq. \eqref{eq:Edyn_Cdyn_one_mode}, which is in good agreement with the numerics shown in Fig.~\ref{fig:Edyn_Cdyn_ED} based on ED for all tested velocities, demonstrating the connection between the breakdown of the dynamo effect and the change of topology.

\subsection{Continuous bath case}

For a spin coupled to an Ohmic bath, effects of the bath on the dynamically measured topology of the spin have been discussed in \cite{henriet2017topology} and are manifest in the results shown in Fig. \ref{fig:numerics_comparison}. Notably, for more pronounced coupling ($\alpha = 0.2$ in the right column of Fig. \ref{fig:numerics_comparison}) we observed a ``bump'' around time $t = \pi/(2v)$ in the dynamics of $\moy{\sigma^y(t)}$. From Eq.~\eqref{eq:Cdyn_sy_integral}, $C_\text{dyn}$ is given in terms of an integral of $\moy{\sigma^y}$ from north to south pole. This ``bump'' is a manifestation of the concentration of Berry curvature around the equator for increasing coupling strength \cite{henriet2017topology}.

To see the relation between the dynamically measured topology of the spin and the dynamo effect,
note that for an Ohmic bath, we can express the dynamically induced field in terms of $\moy{\dot{\sigma}^z}$ using Eq. \eqref{eq:h_from_zt_dyn}. Therefore, we can immediately relate
\bb
\int_0^{\pi/v} dt' h_\text{dyn}(t') =  4 \alpha \pi C_\text{dyn},
\ee
so that the dynamically induced field integrated for a sweep from the north to the south pole is proportional to the dynamically measured Chern number.\\

\begin{figure}
    \centering
    \includegraphics[width = 0.45\textwidth]{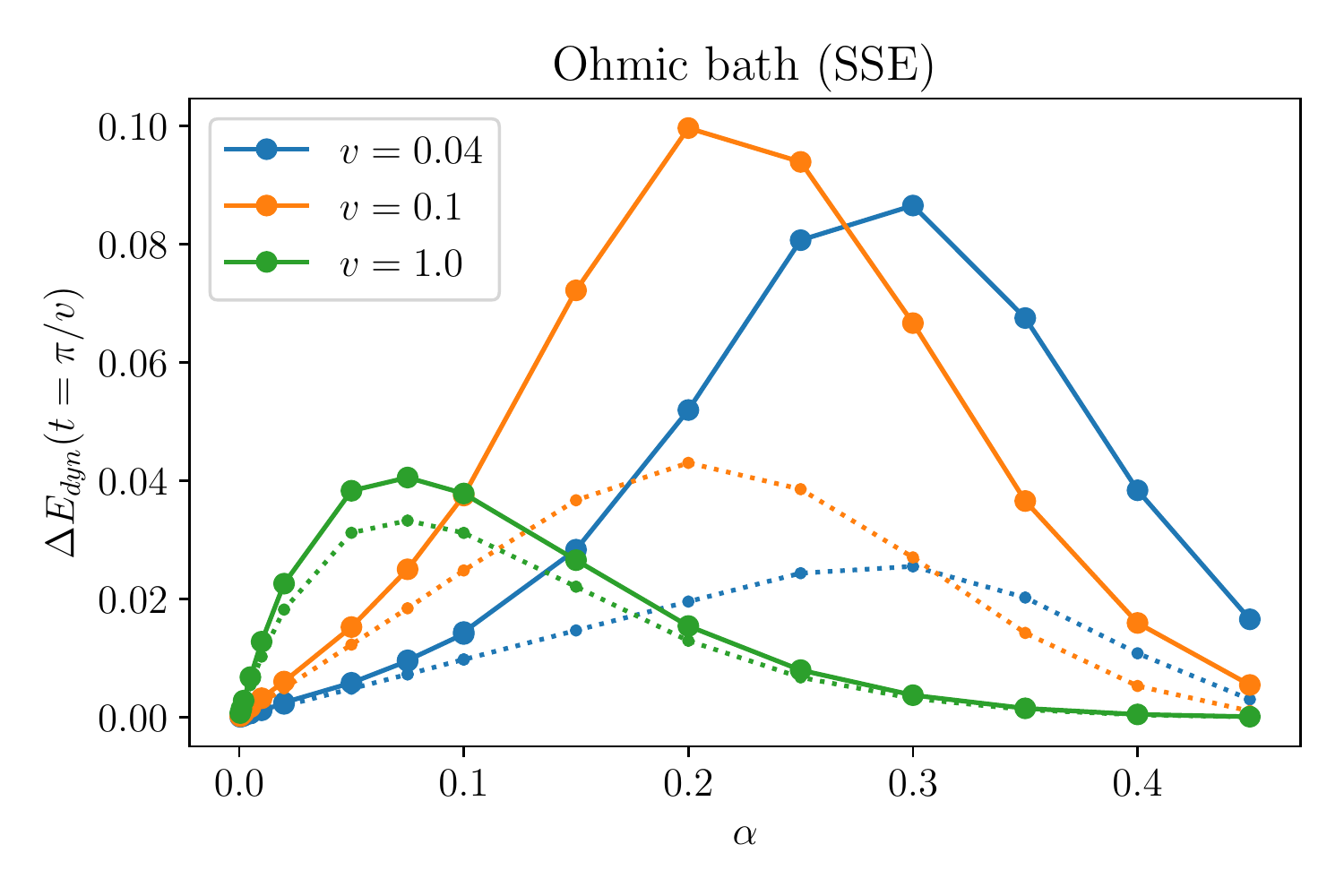}
    \caption{The dynamo energy for coupling to a continuous bath after a drive from north to south pole from the stochastic approach (dots connected by solid lines) and from Eq.~\eqref{eq:free_Edyn_Cdyn} with $C_\text{dyn}$ evaluated from the same results of the stochastic approach (dots connected by dotted lines). Here, $H=1.0$ and $\omega_\text{c} = 100.0$. }
    \label{fig:SSE_Edyn}
\end{figure}

Note that assuming the adiabatic free spin dynamics $\moy{\sigma^z(t)} \sim \cos(v t)$ and using Eq.~\eqref{eq:h_from_zt_dyn} for the dynamically induced field, we can evaluate the dynamo energy $\Delta E_\text{dyn}$ accumulated between $t=0$ and $t=\pi/v$  from Eq.~\eqref{eq:E_dyn_derivative} and find
\bb
    \label{eq:free_Edyn}
 \Delta E_\text{dyn} &=& \frac{\alpha \pi^2 v}{4},
\ee
which corresponds to the non-dissipative limit of Eq.~\eqref{eq:periodic_Edyn}.

Numerical results from the SSE approach are provided in Fig.~\ref{fig:SSE_Edyn}. At small $\alpha\ll 1$, they suggest an extension of \eqref{eq:free_Edyn} as
\bb
\label{eq:free_Edyn_Cdyn}
\Delta E_\text{dyn} &\sim& \frac{\alpha \pi^2 v}{4} C_\text{dyn}^2.
\ee

For larger $\alpha$, this formula does not give the correct result anymore, but interestingly the breakdown of the dynamo effect is still related to $C_\text{dyn}^2$. Similarly, while at weak coupling the dynamo energy scales linearly in the velocity, for stronger couplings there is a trade-off between the linear factor of $v$ in Eq.~\eqref{eq:free_Edyn_Cdyn} and the square of the dynamically measured Chern number which decreases from $C_\text{dyn}=1$ for increasing velocities. From Fig.~\ref{fig:SSE_Edyn} we see that at intermediate coupling strengths, the numerical results deviate from the prediction in Eq.~\eqref{eq:free_Edyn_Cdyn} which does not describe the maximal dynamo energy correctly.

\begin{figure*}
    \centering
    \includegraphics[width = \textwidth]{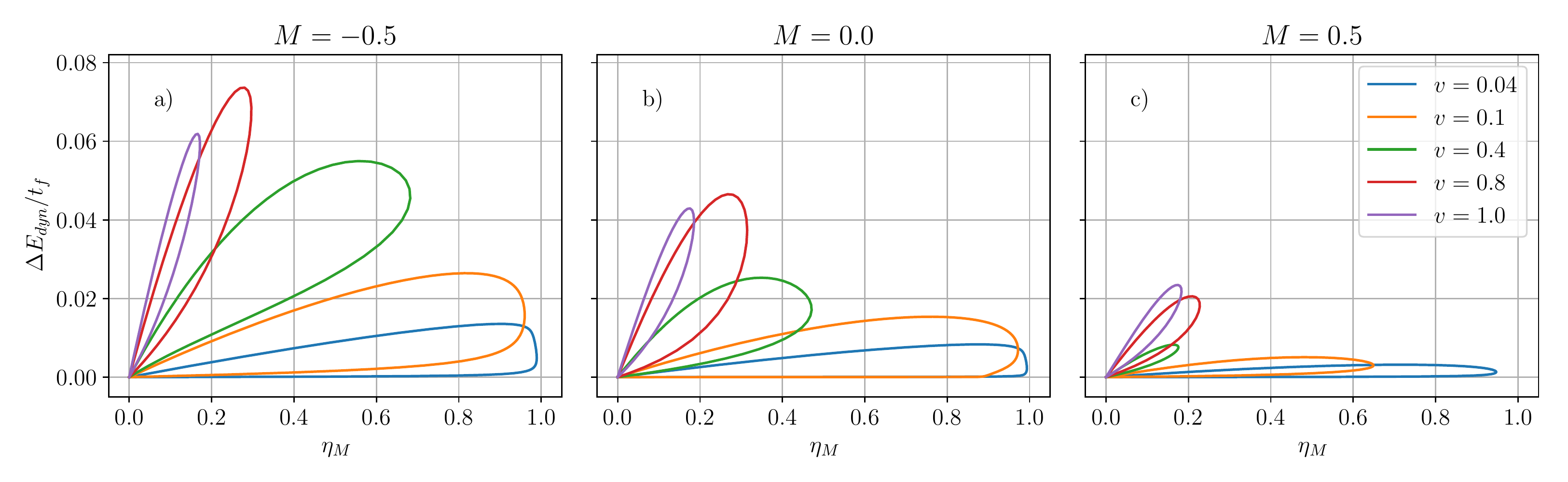}
    \caption{The averaged output power for coupling to one mode as a function of the efficiency $\eta_M$ defined in Eq.~\eqref{eq:eta_M} with a parameter $g/v$ from ED simulation with a) $M = -0.5$, b) $M=0.0$ and c) $M=0.5$. The system is initialized in preparation (1) and with $H=1.0$. The Hilbert space of the bosonic mode is truncated for each data point at a value well beyond the expectation value of the filling and the parameter $g/v$ ranges from $0.0$ to $8.0$.}
    \label{fig:M_dynamo}
\end{figure*}

The observed behavior at weak coupling $\alpha\ll 1$ can be understood using the periodic orbit predicted by the GKLS master equation, from which the  dynamical Chern number can be calculated straightforwardly to be $C_\text{dyn} = H/\Omega$. The dynamo energy accumulated during the half rotation reads from Eq. \eqref{eq:periodic_Edyn}:
\bb
\Delta E_\text{dyn} = \frac{\alpha \pi^2 v}{4} C_\text{dyn}^2,
\ee
showing exact proportionality to $C_\text{dyn}$ within the validity regime of the GKLS master equation, which includes in particular non-adiabatic driving speeds.

Interestingly, we see a trade-off between topological protection of the dynamo effect, which occurs for $v\ll H$ such that $C_\text{dyn}\to C = 1$, while the dynamo energy goes to zero for small speed.

\subsection{Adding a constant bias field}

So far, we have considered $M=0$ in Eq. \eqref{eq:spin_radial_field}. For the resonant one mode dynamo, we argued that its role is in some sense played by the dynamically induced field: When it is small, it has little influence on the spin dynamics and leaves $C_\text{dyn}$ unchanged. However, when it grows too big, at a certain magnitude it fixes the spin and thereby alters the topology of the dynamically accessed effective ground state manifold of the spin which is reflected in $C_\text{dyn} \to 0$ and a simultaneous breakdown of the dynamo effect. This has the curious implication that a bias field which opposes the effect of the dynamically induced field may stabilize the dynamo effect by preventing its breakdown. Explicitly, we are considering preparation (1) where initially the induced field is $h(0) = -g^2/v$. Therefore, including $0>M>-H$, the extenal bias field opposes the induced field while the prepared spin state $\ket{\uparrow}$ is still the ground state. It is then possible to induce larger fields before the dynamo effect breaks down. We study this system numerically by ED simulations and find that the maximal output power when driving the system from $t=0$ to $t=\pi/v$ is increased with these parameters. To compute the efficiency in this setup, we substract the introduced energy imbalance $M$ between the spin ground states at the north and the south pole from the work, i.e.
\bb
\label{eq:eta_M}
\eta_M = \frac{\Delta E_\text{dyn}}{W_\text{dr} - M}
\ee
This definition reflects the fact that $M$ is a one-shot resource that can be consumed after initialization. Note that the definition of $\eta$ in Eq. \eqref{eq:efficiency} is still sensible over full periods of the driving.
The numerical results are shown in Fig. \ref{fig:M_dynamo}. We see by comparing Fig. \ref{fig:M_dynamo} a) to Fig. \ref{fig:M_dynamo} b) that higher output powers can be achieved while high efficiencies are still possible.
In this way, adding a constant negative offset field $M$ constitutes a way to adapt the dynamo to desired output parameters. To the contrary, adding a positive field decreases the maximum output power, as seen in Fig.~\ref{fig:M_dynamo} c).
The dynamo energy is still related to the dynamically measured Chern number $C_\text{dyn}$ through Eq.~\eqref{eq:free_Edyn_Cdyn}, so this way to modify the dynamo is inherently possible because of the topological properties of the considered system.

\section{Conclusions}\label{sec:conclusions}

In this article, we discussed the quantum dynamo effect occurring in an externally driven system coupled to a bosonic environment.
We have demonstrated how the driving can induce a coherent field in the environment, which allows to interpret the energy transferred to the bosonic modes as work.
By focusing on a periodically driven spin-$\frac{1}{2}$, we exemplified the mechanism of the quantum dynamo and its performance for the cases of coupling to only one mode and to an Ohmic bath with a high-frecuency cut-off. These setups correspond to realizations where the spin is coupled to a cavity mode \cite{viennot2015coherent} or to a long transmission line \cite{cedraschi2001quantum,forn2017ultrastrong,magazzu2018probing} respectively.\par
From the one-mode scenario, we demonstrated that the effect occurs for a mode with a frequency close to resonance with respect to the external driving velocity.
In the limits of weak and strong coupling to the bath (corresponding respectively to the spin following the external drive or being polarized in its initial direction), we provided an analytical understanding of the effect, of the field induced due to the dynamical response of the resonant mode and of the amount of energy transferred to the mode. Using exact diagonalization, we verified these results numerically and analyzed the performance of the dynamo.
We observed that the energy conversion can reach efficiencies close to unity at small driving velocities, while higher velocities allow for larger output powers averaged over the time of the driving.
We demonstrated that the single-mode dynamo breaks downs when a large field has been induced, such that the polarized spin limit is eventually reached.
We then showed that coupling the spin to a finite number of modes, following a discretized Ohmic spectral density, solves this issue. Using ED simulations, we showed that the dynamo effect still occurs for modes around the resonant frequency, whereas the other modes are useful to increase the dynamo stability over several periods.\par
We finally considered the full problem of a continuous Ohmic bosonic bath where we employed a numerically exact stochastic Schr\"odinger equation approach to simulate the spin evolution under the influence of the bath. To describe the dynamics in the long time limit, we developed a GKLS master equation approach, demonstrating that the spin reaches a periodic orbit defined by the joint influence of driving and dissipation. Using analytical arguments to evaluate the various contributions to the induced field in the bath, we demonstrated how they can be evaluated from the observables of the spin only. We then discussed the performances of the dynamo from the numerical results of the stochastic Schr\"odinger equation approach and the GKLS master equation. The latter allowed us to predict an efficiency approaching unity at weak coupling in the long-time limit, confirming the trend that increasing the number of bosonic modes stabilizes the dynamo.\par
Finally, we related our results to the topological properties of the driven spin. From the numerical results and based on analytical arguments from limiting cases, we argued that the energy converted into coherent displacement of the bath modes due to the dynamo effect is related to the square of the dynamically measured Chern number. This lead us to the conclusion that at adiabatic driving velocities, the effect enjoys topological protection. By adding an external bias field, we extended this analysis and in the same time provided a way to tune the output power of the effect.\par
Such effects can be verified experimentally, e.g. in superconducting circuits where the driven artificial atoms can be coupled to microwave cavities or transmission lines, and the emitted field can be characterized with great accuracy \cite{Peronnin20}.

\begin{acknowledgments}
We thank Lo\"ic Henriet, Peter Orth, Michel Ferrero and Alberto Rosso for interesting discussions related to this work.
EB \& KLH acknowledge support from the Deutsche Forschungsgemeinschaft (DFG) under project number 277974659.
CE acknowledges funding from the French National Research Agency (ANR) under grant ANR-20-ERC9-0010 (project QSTEAM).
KLH also acknowledges support from ANR BOCA.
\end{acknowledgments}

\onecolumngrid
\appendix
\section{Path integral formulation of the spin dynamics}\label{app:path_integral}
As mentioned in Sec. \ref{sec:multimode_general}, the influence of a harmonic bath on the spin in the formulation of the Hamiltonian in Eq. \eqref{d:model} can be encoded in a double path integral over all spin paths of the influence functional \cite{feynman1963theory}. To explain what insights can be drawn from this formulation, we are mainly following Refs. \cite{orth2013nonperturbative, loicphd} in this appendix.
Throughout this work we consider the system to be prepared such that the spin is initially pointing upward in $z$-direction. Transitions (spin flips) are induced by the transverse field and the coupling to the bath contributes to the weight. It can be shown that the elements of the spin reduced density matrix can be expressed as \cite{feynman1963theory, leggett1987dynamics}
\begin{equation}
  \label{eq:spin_path_integral_app}
  \bra{\sigma_f} \rho_s(t) \ket{\sigma'_f} = \int \mathcal{D}\sigma \int \mathcal{D}\sigma' \mathcal{A}[\sigma]\mathcal{A^*}[\sigma'] F[\sigma,\sigma'].
\end{equation}
The functionals $\mathcal{A}[\sigma]$ occurring in Eq.~\eqref{eq:spin_path_integral_app} contribute as the amplitude of a path when the spin evolves freely and thus are formally given in terms of the free (i.e. without coupling to the bath) action of the spin $\mathcal{A}[\sigma] = \exp(i S_\text{free})$ \cite{weiss2012quantum, feynman1963theory}. In the real-time path integral formulation of Eq.~\eqref{eq:spin_path_integral_app}, $\sigma^z$ gives rise to a weight factor depending on $\sigma$, while $\sigma^x$ induces spin flips from $\sigma =\pm 1$ to $\sigma = \mp 1 $. It therefore allows to rewrite the path integral as a series of spin flips \cite{orth2013nonperturbative, loicphd}:
\bb
 \int \mathcal{D}\sigma \mathcal{A}[\sigma] &=& \sum_{n=0}^\infty \int_0^t dt_n \frac{-i H \sin(vt_n)}{2} \int_0^{t_n} dt_{n-1} \frac{-i H \sin(vt_{n-1})}{2} \times \cdots \nonumber\\
 &\cdots& \times \int_{0}^{t_2} dt_1 \frac{-i H \sin(vt_{1})}{2} \times \exp \left(-i \int_0^t d\tau \frac{H\cos(v\tau)}{2} \sigma(\tau)\mid_{\{t_1, t_2,\ldots,t_n \}} \right), \label{eq:DsigmaAsigma}
\ee
where we denoted a classical spin path with $n$ spin flips at times $t_1, t_2,\ldots,t_n$ by $\sigma(t)\mid_{\{t_1, t_2,\ldots,t_n \}}$.
The influence functional $F$ encodes the influence of coupling to the environment and is given by \cite{feynman1963theory, leggett1987dynamics, orth2013nonperturbative}
\bb
    \label{eq:influence_functional}
  F[\sigma,\sigma'] = \exp \bigg(-\frac{1}{\pi} \int_{t_0}^{t} ds \int_{t_0}^{s} ds' [-iL_1(s-s')\xi(s) \eta(s')+L_2(s-s') \xi(s)\xi(s')] \bigg),
\ee
where we introduced the reparametrization $\eta(t) = \frac{1}{2}(\sigma(t) + \sigma'(t))$ and $\xi(t) = \frac{1}{2}(\sigma(t) - \sigma'(t))$ representing the symmetric and antisymmetric spin paths and
\begin{align}
  L_1(t) &= \int_0^\infty d\omega J(\omega) \sin(\omega t),\\
  L_2(t) &= \int_0^\infty d\omega J(\omega) \cos(\omega t) \coth(\frac{\beta \omega}{2}),
\end{align}
are the bath correlation functions defined from \cite{orth2013nonperturbative}
\begin{equation}
  \pi  \moy{X(t)X(0)} = L_2(t) - i L_1(t),
\end{equation}
where $X(t) = \sum_k g_k (b_k^\dag + b_k)$.
We first consider paths from an initial state $\ket{\uparrow}$ to a final state $\ket{\uparrow}$. This corresponds to the upper left element in the density matrix. Diagonal elements in this framework are called `sojourn states', for which both of the classical variables $\sigma$ and $\sigma'$ take the same value, while states where $\sigma$ and $\sigma'$ have opposite values are called `blip states'. Adapting a notation in which we summarize the double path in the configuration space for one spin to a single path in configuration space for both $\sigma$ and $\sigma'$, the paths $\eta$ and $\xi$ can be parametrized by the times at which a spin is flipped along the path,
\begin{align}
  \xi(t) &= \sum_{j=1}^{2n} \Xi_j \theta(t-t_j),\\
  \eta(t) &= \sum_{j=0}^{2n} \Upsilon_j \theta(t-t_j),
\end{align}
with $\{\Xi_1, \Xi_2,...,\Xi_{2n}\} = \{\xi_1, -\xi_1,...,-\xi_{n}\}$
and $\{\Upsilon_1, \Upsilon_2,...,\Upsilon_{2n} \} = \{\eta_0,-\eta_0,... \eta_n \}$
and using this parametrization, the influence functional can be simplified:
\begin{equation}\label{influence_functional}
  F_n [\{ \Xi_j \},\{ \Upsilon_j \},\{ t_j \}] = \mathcal{Q}_1 \mathcal{Q}_2,
\end{equation}
with
\begin{align}
  \mathcal{Q}_1 &= \exp( \frac{i}{\pi} \sum_{j > k\geq 0}^{2n} \Xi_j \Upsilon_k Q_1(t_j-t_k) ), \label{curl_Q1}\\
  \mathcal{Q}_2 &= \exp( \frac{1}{\pi} \sum_{j > k\geq 1}^{2n} \Xi_j \Xi_k Q_2 (t_j-t_k) ). \label{curl_Q2}
\end{align}
Here, $Q_{1}(t)$ and $Q_2(t)$ are the second integrals in time of $L_1(t)$ and $L_2(t)$ respectively. For an Ohmic spectral density $J(\omega) = 2\pi\alpha \omega e^{-\omega/\omega_\text{c}}$ and $T=0$ we can evaluate specifically
\begin{align}
  Q_1(t) &= 2\pi\alpha \arctan(\omega_\text{c} t),\label{Q1_equation} \\
  Q_2(t) &= \pi \alpha \log(1 + \omega_\text{c}^2 t^2).
\end{align}
Focusing on the upper left element of the density matrix, we denote it by $p(t) = \bra{\uparrow} \rho_s(t) \ket{\uparrow} = \frac{1+\langle \sigma^z(t) \rangle}{2}$. For a path from up to up we need to have an even number of spin flips and $\Upsilon_0 = \Upsilon_{2n} = 1$. Using this, rewriting the series expansion coming from $\int \mathcal{D}\sigma \mathcal{A}[\sigma]$ described in Eq.\eqref{eq:DsigmaAsigma} together with a similar expression for $\int \mathcal{D}\sigma' \mathcal{A}^*[\sigma']$ in terms of the symmetric and the antisymmetric spin paths $\eta(t)$ and $\xi(t)$ and finally plugging in Eq.~\eqref{influence_functional}, we find \cite{orth2013nonperturbative}:
\bb
\label{eq:matrix_element_up_up}
  p(t) = 1 + \sum_{n=1}^\infty \int_0^t dt_{2n} \frac{iH\sin(vt_{2n})}{2}\times...\times \int_0^{t_2} dt_1 \frac{iH\sin(vt_1)}{2}\sum_{(\Xi_j, \Upsilon_j)} F_n H_n.
\ee
The factor $H_n$ comes from the contribution of the bias field to the amplitudes $\mathcal{A}[\sigma]$ and is defined by
\begin{equation}\label{deterministic_H}
  H_n = \exp (-i \sum_{j=1}^{2n} \Xi_j \int_0^{t_j} H\cos(vt') dt').
\end{equation}
Note that in Eq.~\eqref{eq:matrix_element_up_up}, a blip state is coupled to all previous blip and sojourn states through the influence functional, thus hindering an analytical solution. Therefore, in the following we first present a well-known approximation scheme known as `NIBA' \cite{leggett1987dynamics, dekker1987noninteracting, dakhnovskii1994dynamics}. After that we show how a stochastic evaluation of the influence functional leads to a numerically exact approach to solve the spin dynamics.

\subsection{NIBA approximation}
\label{app:NIBA}

Neglecting the interblip correlations and blip-sojourn correlations except for neighbouring ones leads to the following expressions for the spin expectation values \cite{grifoni1998driven, grifoni1996exact}:
\begin{subequations}\label{eq:NIBA}
  \begin{align}
    \moy{\dot{\sigma}^z(t)} &= \int_0^t dt' \left(K^{(-)}(t,t') - K^{(+)}(t,t') \moy{\sigma^z(t')} \right),\\
    \moy{\sigma^x(t)} &= \int_0^t dt' \left(Y^{(+)}(t,t') + Y^{(-)}(t,t') \moy{\sigma^z(t')} \right), \label{eq:NIBA_x} \\
    \moy{\sigma^y(t)} &= -\frac{1}{H \sin(vt)}\moy{\dot{\sigma}^z(t)},
\end{align}
\end{subequations}

with
\begin{subequations}
  \begin{align}
    K^{(+)}(t,t') &= \Delta(t) \Delta(t') e^{-Q_2(t-t')} \cos(Q_1(t-t')) \cos(\zeta(t,t')),\\
    K^{(-)}(t,t') &= \Delta(t) \Delta(t') e^{-Q_2(t-t')} \sin(Q_1(t-t')) \sin(\zeta(t,t')),
\end{align}
\end{subequations}
and
\begin{subequations}
  \begin{align}
    Y^{(+)}(t,t') &= \Delta(t) \Delta(t') e^{-Q_2(t-t')} \sin(Q_1(t-t')) \cos(\zeta(t,t')),\\
    Y^{(-)}(t,t') &= \Delta(t) \Delta(t') e^{-Q_2(t-t')} \cos(Q_1(t-t')) \sin(\zeta(t,t')),
\end{align}
\end{subequations}
where $\Delta(t) = H\sin(vt)$ and $\zeta(t,t')=\int_{t'}^t dt'' H\cos(vt'')$ for our realization with Hamiltonian \eqref{d:model}. Eqs. \eqref{eq:NIBA} can be solved numerically, which is shown in Fig. \ref{fig:numerics_comparison} by the dotted lines.

\subsection{Stochastic approach}\label{app:SSE}

To find a result for the spin dynamics which is numerically exact, we can tackle the problem of calculating the double path integral in Eq. \eqref{eq:spin_path_integral_app} stochastically following
\cite{orth2013nonperturbative, loicphd}.
We want to determine $p(t)$ defined in Eq.~\eqref{eq:matrix_element_up_up}, which can be realized by evaluating $F_n$ in a stochastic way. To simplify, note that for large $\omega_\text{c}$, one can take $Q_1(t) \approx \pi^2 \alpha$ \cite{orth2013nonperturbative}. Furthermore, $\mathcal{Q}_2$ can be expressed in terms of a stochastic field satisfying \cite{loicphd}
\begin{subequations}\label{h_s}
  \begin{align}
  \overline{ h_s(t) } &= 0,\\
  \overline{h_s(t)h_s(s)} &= \frac{1}{\pi} Q_2(t-s)+l_1,
  \end{align}
\end{subequations}
where $l_1$ is a complex constant. This allows to rewrite Eq. \eqref{influence_functional} through \eqref{curl_Q2}:
\begin{align}
  F_n &= \mathcal{Q}_1 \mathcal{Q}_2, \\
  &= \exp( i \alpha \pi \sum_{j > k\geq 0}^{2n} \Xi_j \Upsilon_k  ) \exp( \sum_{j > k\geq 1}^{2n} \Xi_j \Xi_k (\overline{h(t_j)h(t_k)}-l_1) ),\\
  &= \exp( i \alpha \pi \sum_{k = 0}^{n-1} \xi_{k+1} \eta_k ) \overline{\exp(\sum_{j=1}^{2n}h_s(t_j)\Xi_j)}. \label{Fn_average}
\end{align}
Note that to go beyond the approximation $Q_1(t) \approx \pi^2 \alpha$, one can introduce a second stochastic field capturing $Q_1(t)$ fully \cite{henriet2014quantum}. In this article, we worked with only one stochastic field as we assumed $\omega_c$ to be the biggest energy scale in the model. We checked that the spin dynamics thus achieved agrees with analytical results, see Fig.~\ref{fig:bethe} in the main text.\par
The expression for $p(t)$ can be rewritten using a time-ordered exponential \cite{orth2013nonperturbative}:
\begin{equation}
  p(t) = \overline{\bra{\Phi_f} \mathcal{T} e^{-i \int_0^t dt' V(t')}  \ket{\Phi_i}}.
\end{equation}
 One thus finds a stochastic Schr\"odinger equation (SSE)
 \begin{equation}\label{eq:SSE}
  i \partial_t \ket{\Phi} = V(t) \ket{\Phi},
\end{equation}
which can be solved to evaluate one realization of $p(t)$.
From Eq. \eqref{Fn_average}, we can write $V(t)$ as \cite{loicphd}
\begin{equation}
  V(t) = \frac{H \sin(vt)}{2} \begin{pmatrix}
        0 & e^{-h} & -e^{h} & 0 \\
        e^{i \pi \alpha} e^{h} & 0 & 0 & -e^{-i \pi \alpha} e^{h}\\
        -e^{-i \pi \alpha} e^{-h} & 0 & 0 & e^{i \pi \alpha} e^{-h}\\
        0 & -e^{-h} & e^{h} & 0
      \end{pmatrix},
\end{equation}
and $h(t)$ is defined from Eq. \eqref{deterministic_H} and $h_s(t)$ by
\begin{equation}
  h(t) = h_s - i \int_0^t H\cos(vt') dt'.
\end{equation}
Averaging over many realizations of the stochastic field, one finds the spin expectation value.
To realize the field $h_s(t)$ with the properties \eqref{h_s}, we decompose $Q_2(t)$ in a Fourier series. For this, realize that $Q_2$ is a symmetric function. We can define $\tau = t/t_f$ so that $Q_2(\tau t_f)$ is defined on $\tau \in [ -1,1 ] $. If we extend this to make $Q_2(\tau t_f)$ a periodic function, it is justified to write it as a Fourier series.
We can write \cite{loicphd}
\begin{align}
  \frac{Q_2 (\tau t_f)}{\pi} &= \frac{g_0}{2} + \sum_{m=1}^{\infty} g_m \cos(m\pi\tau),
\end{align}
because of the symmetry. Then we can define $h_s(t)$ in terms of the Fourier coefficients as
\begin{align}
  h_s(\tau t_f) = \sum_{m=1}^\infty (g_m)^{1/2}\left[ s_{1,m} \cos(\pi m \tau) - s_{2,m} \sin(\pi m \tau) \right],
\end{align}
which with $s_{1,m}, s_{2,m}$ being independent normal Gaussian variables gives the properties \eqref{h_s}. Finally, this realization of the stochastic field is used to evaluate $V(t)$ from which the SSE \eqref{eq:SSE} is solved. The result is then the solution averaged over the realizations.
In the simulation, the Fourier coefficients are found using the fast Fourier transform.\par
Finally, to find expectation values of $\sigma^{x/y}$, we need to evaluate off-diagonal elements of the density matrix. Those have contributions from paths which make $2n-1$ transitions. Therefore, we need to modify the parametrization of the paths to account for that and modify the expression of the influence functional such that paths end after $2n-1$ transitions in a blip state. Essentially, it is as if the system stepped back one time step from a final sojourn state, therefore in order to find $[\rho_s(t)]_{12}$, we need to project out the second component multiplied by $e^{-h(t)}$.
Eventually, it holds that $[\rho_s(t)]_{ij} = \overline{\bra{\Sigma_{ij}} \ket{\Phi(t)}} $ where $\bra{\Sigma_{11}} = (1,0,0,0)$, $\bra{\Sigma_{12}} = (0,e^{-h(t)},0,0)$, $\bra{\Sigma_{21}} = (0,0,e^{h(t)},0)$ and $\bra{\Sigma_{22}} = (0,0,0,1)$ \cite{loicphd}.

\subsection{Comparison of results from the stochastic approach to ED}\label{app:hard_cutoff}

\begin{figure}[ht]
    \centering
     \includegraphics[width=\textwidth]{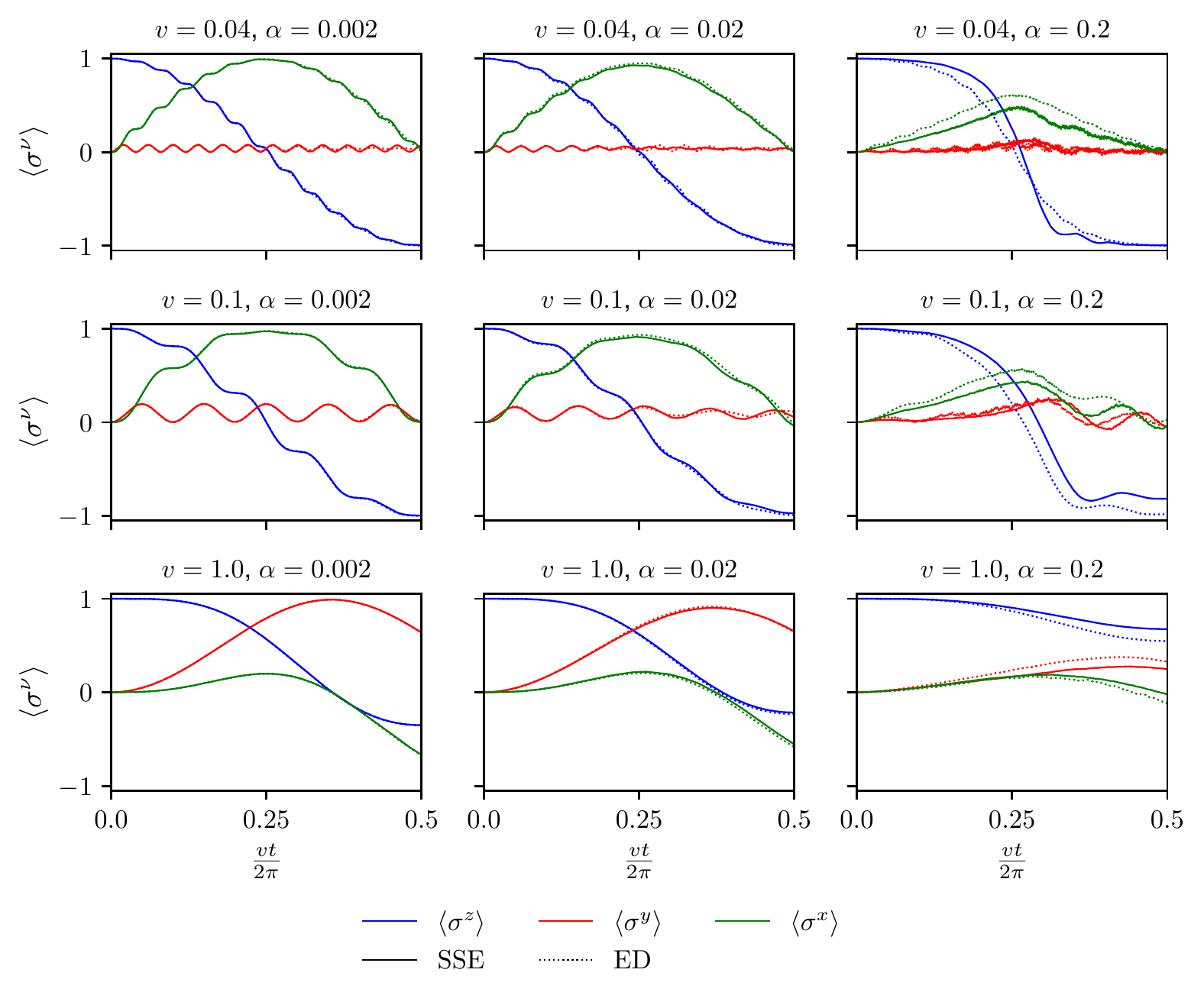}
        \caption{Comparison of numerical results between ED (dotted line) and stochastic approach (solid line). The expectation values of $\moy{\sigma^z}$, $\moy{\sigma^y}$ and $\moy{\sigma^x}$ are shown in blue, green and red in this order. Here, $H = 1.0$ and $\omega_\text{c} = 100.0$. For the ED result, we discretized the spectrum with a hard cut-off at $\omega_\text{c}$ into ten modes. Their Hilbert space was truncated such that the maximal occupation is well below the truncation. }

        \label{fig:SSE_ED_comparison}
\end{figure}

To demonstrate the validity of the results obtained using the stochastic approach, we show a comparison to results obtained using exact diagonalization with a finite number of modes as lined out in Sec. \ref{sec:finite_number_modes}. The numerical results are shown in Fig. \ref{fig:SSE_ED_comparison}. We see that for all shown velocities, the ED agrees very well with the stochastic approach for weak coupling. At stronger coupling, there are deviations between the two but still there is qualitative agreement. We therefore conclude that the results from the stochastic approach in the considered range of parameters correctly reproduce the physical situation.\par
Note that for the ED, we used a discretization of the spectral function of the form of Eq.~\eqref{eq:hard_cutoff_spectral_density} with a hard cut-off at $\omega_\text{c}$, while for the SSE the cut-off is implemented in an exponential form (cf. Eq.~\eqref{eq:ohmic_spectral_density}). Note that with a hard cut-off, the Kernel function defined in Eq.~\eqref{eq:kernel} to evaluate the induced field from the system evolution is modified:
\bb
K(t) &=& -2 \sum_k g_k^2 \sin(\omega_k t) = -2 \int_0^{\infty} \frac{J(\omega)}{\pi} \sin(\omega t) = -4 \alpha \int_0^{\omega_\text{c}} \! d\omega \, \omega \sin(\omega t),\nonumber \\
&=& -4 \alpha \frac{\omega_\text{c}^2 \sin (\omega_\text{c} t)-\omega_\text{c}^2 (\omega_\text{c} t) \cos (\omega_\text{c} t)}{(\omega_\text{c} t)^2}.
\ee
On the other hand, using the approximation $\moy{\sigma^z(t)}\sim \cos(vt)$, we can evaluate the induced field of each mode from Eq. \eqref{h_onemode_cos_1} and sum over the modes, i.e. (in preparation (1))
\bb
h(t) &=& \sum_k h_{k} = \sum_k \frac{g_k^2 \omega_k}{\omega_k^2-v^2} \left(\frac{v^2}{\omega_k^2}\cos(\omega_k t) - \cos(vt) \right)= 2 \alpha \int_0^{\omega_\text{c}} d\omega  \frac{1}{\omega^2-v^2} \left(v^2\cos(\omega t) - \omega^2\cos(vt) \right).
\ee
Decomposing the double pole into a sum of two simple poles, we can evaluate for $t>0$ and $\omega_\text{c} t \gg 1$
\bb
h(t) &=&  -\alpha \cos(vt) \bigg(2\omega_\text{c} + v\log(\frac{\omega_\text{c}-v}{\omega_\text{c}+v}) \bigg) - \alpha \pi v \sin(vt) \approx -2 \alpha \omega_\text{c} \cos(vt) - \alpha \pi v \sin(v t),
\ee
where the last step is justified if $\omega_\text{c} \gg v$. This result is consistent with Eqs.~\eqref{eq:h_from_zt_ad} and \eqref{eq:h_from_zt_dyn} so we conclude that in this limit the induced field is the same for an Ohmic spectral function with an exponential and a hard cut-off in the continuum limit.

\section{Derivation of the GKLS equation}\label{app:GKLS}
In this appendix, we provide the derivation of the Floquet-Markov GKLS master equation (Eq.~\eqref{eq:GKLS} of the main text), following \cite{cohentannoudjibook,breuerbook}.
We start from the exact spin-reservoir dynamics given by:
\bb
\dot\rho_{SR}(t) = -i[{\cal H}(t),\rho_{SR}].
\ee
So as to remove the time-dependence in the spin Hamiltonian, we first rewrite the dynamics into a rotating frame defined by the unitary transformation $U_\text{rot} = e^{it\frac{v}{2}\sigma_y}$:
\bb
\dot\rho_\text{rot} = -i[H_\text{rot}(t), \rho_\text{rot}(t)],
\ee
where $\rho_\text{rot} = U_\text{rot}\rho_{SR}U_\text{rot}^\dagger$ and the relevant Hamiltonian is:
\bb
\tilde H_\text{rot}(t) &=& -\frac{H}{2}\sigma_z - \frac{v}{2}\sigma_y +  (\cos(vt)\sigma_z -\sin(vt)\sigma_x)R + {\cal H}_R.
\ee
In this frame, the spin dynamics is ruled by ${\cal H}_S =  -\frac{H}{2}\sigma_z - \frac{v}{2}\sigma_y $ of eigenvalues $\pm\Omega/2$, where we have introduced $\Omega = \sqrt{H^2+v^2}$, associated with eigenstates $\{\ket{+},\ket{-}\}$ defined by:

\bb
\ket{+} &=& i\sqrt{\frac{\Omega-H}{2\Omega}}\ket{\uparrow} + \sqrt{\frac{\Omega+H}{2\Omega}}\ket{\downarrow}\\
\ket{-} &=& \sqrt{\frac{\Omega+H}{2\Omega}}\ket{\uparrow} + i\sqrt{\frac{\Omega-H}{2\Omega}}\ket{\downarrow}.
\ee
When written back in the lab frame, $U_\text{rot}^\dagger(t)\ket{\pm}$ correspond to the Floquet states which in this case can be computed analytically. On the other hand, the qubit operator coupled to the bath is now explicitly time-dependent. We then apply a second unitary $U_{I} = e^{i{\cal H}_St}$ to go to the interaction picture with respect to the free spin dynamics, that we denote with a tilde. We obtain:
\bb
\dot{\tilde\rho}_{SR}(t) = -i[ \tilde{\cal H}(t), \tilde\rho_{SR}(t)],\label{eq:LiouvI}
\ee
where
\bb
\tilde{\cal H}(t) = \sum_{l = \pm}\sum_{\omega = \pm \Omega,0} \hat s_{l,\omega} \tilde R(t) e^{i(\omega+ l v)t},
\ee
in terms of the operators:
\bb
\hat s_{l,\Omega} &=&  -\frac{i}{2}\frac{v+l\Omega}{\Omega}\ket{+}\bra{-},\\
\hat s_{l,-\Omega} &=&  \frac{i}{2}\frac{v-l\Omega}{\Omega}\ket{-}\bra{+},\\
\hat s_{l,0} &=& -\frac{H}{2\Omega}(\ket{+}\bra{+}-\ket{-}\bra{-}).
\ee
We are now ready to apply the standard derivation of the GKLS equation \cite{cohentannoudjibook}. To take advantage of the time-scale separation between the reservoir internal dynamics and the system relaxation, we coarse-grain this equation over a time-scale $\Delta t$ and take the trace over the reservoir subspace, defining $\Delta\tilde\rho_S = \text{Tr}_R\{\tilde\rho_{SR}(t+\Delta t)-\tilde\rho_{SR}(t)\}$:
\bb
\frac{\Delta \tilde\rho_{S}}{\Delta t} &=& \frac{-i}{\Delta t}\int_0^{\Delta t} dt' \text{Tr}_R[\tilde{\cal H}(t'),\tilde\rho_{SR}(t')],\nonumber\\
&=& \frac{-i}{\Delta t}\int_0^{\Delta t} dt' \text{Tr}_R[\tilde{\cal H}(t'),\tilde\rho_{SR}(0)]-\frac{1}{\Delta t}\int_0^{\Delta t}dt' \int_0^{t'} dt'' \text{Tr}_R[\tilde{\cal H}(t'),[\tilde{\cal H}(t''),\tilde\rho_{SR}(t'')].
\ee
To go to the last line, we inserted the formal integral solution of Eq.~\eqref{eq:LiouvI}. We moreover assumed that the initial state of the reservoir is the vacuum such that the first term on the right-hand side is zero. We now approximate the system-reservoir state in the integrand with a factorized state $\tilde\rho_{S}(t'')\otimes\tilde\rho_{R}^\text{eq}$. The trace then forms correlation functions of the reservoir $C(t'-t'') = \moy{R^I(t')R^I(t'')}$. We also apply the Markovian approximation, neglecting the (slow) variation of the interaction picture spin state during $\Delta t$, therefore replacing $\rho_{S}(t'')$ with $\rho_{S}(t)$ in the integrand:
\bb
\frac{\Delta \tilde\rho_{S}}{\Delta t} &=& -\frac{1}{\Delta t}\int_0^{\Delta t}dt' \int_0^{t'} dt'' C(t'-t'')\sum_{l,l',\omega,\omega'}[ \hat s_{l,\omega}^\dagger, \hat s_{l',\omega'}\tilde\rho_{S}(t)]e^{-i(lv+\omega)t'}e^{i(l'v+\omega')t''} + \text{H.c.},\nonumber\\
 &=& -\frac{1}{\Delta t}\int_0^{\Delta t}dt' \int_0^{t'} d\tau C(\tau)\sum_{l,l',\omega,\omega'}[ \hat s_{l,\omega}^\dagger, \hat s_{l',\omega'}\tilde\rho_{S}(t)]e^{i((l'-l)v+(\omega'-\omega))t'}e^{-i(l'v+\omega')\tau} + \text{H.c.}\label{eq:prec}
\ee
We have:
\bb
 \!\!\int_0^{\infty} \!\! d\tau C(\tau)e^{i \nu \tau} &=& \sum_k g_k^2\left[\moy{b_k^\dagger b_k }\!\!\left(\pi \delta(\omega_k+\nu) +i {\cal P}\frac{1}{\omega_k+\nu}\right) + \moy{b_k^\dagger b_k+1} \!\!\left(\pi \delta(-\omega_k+\nu) -i {\cal P}\frac{1}{\omega_k-\nu}\right) \right],\nonumber\\
&=& \Theta(-\nu)\frac{J(-\nu)}{2} +i\delta_\text{LS}(\nu).
\ee
where we have considered the case of a reservoir in the vacuum (zero temperature) such that $\moy{b_k^\dagger b_k} = 0$ and we have defined $\delta_\text{LS}(\nu) = -i{\cal P}\sum_k \frac{g_k^2}{\omega_k-\nu}$.\\

Eq.~\eqref{eq:prec} is the Bloch-Redfield equation. In the regime $J(\Omega\pm v), J(v) \ll v,\Omega$, requiring in particular $\alpha \ll v/H$, this equation can be further simplified by choosing the coarse-graining time-scale to verify $\Delta t \gg v^{-1},\Omega^{-1}$. As a consequence, all the oscillating terms average out (they give contributions vanishing as $\sin(\omega\Delta t)/(\omega\Delta t)$), leaving only those verifying $l=l'$ and $\omega=\omega'$. This step is often referred to as the secular approximation \cite{cohentannoudjibook}:

\bb
\frac{\Delta \tilde\rho_{S}}{\Delta t} &=& \sum_{l,\omega} \bigg(\Theta(-\omega-lv)\frac{J(-\omega-lv)}{2}\left([ \hat s_{l,\omega}\tilde\rho_{S}(t), \hat s_{l,\omega}^\dagger]+ \text{H.c.}\right) -i  \delta_\text{LS}(\omega+lv)[\hat s_{l,\omega}^\dagger \hat s_{l,\omega},\tilde\rho_{S}(t)]\bigg),\\
&=& -i[\tilde{\cal H}_\text{LS},\rho]+ \sum_l\left( J(\Omega+lv){\cal D}[s_{l,-
\Omega}]\rho + J(v){\cal D}[\hat s_{l,0}]\right)\rho,
\ee
where the Lamb-shift Hamiltonian reads:
\bb
\tilde{\cal H}_\text{LS} = \sum_{l,\omega}  \delta_\text{LS}(\omega+lv)\hat s_{l,\omega}^\dagger \hat s_{l,\omega}.
\ee
Note that the coefficients of this master equation are independent of time in the rotating frame.

Finally, we use that the two operators $\hat s_{\pm,-\Omega}$ are proportional to each other to write it in the form of Eq.~\eqref{eq:GKLS}.

\textit{Relaxation towards the stationary orbit}: In the rotating frame, the state of the spin relaxes towards the pure state $\ket{-}$. Indeed $\hat s_{l,-\Omega}$ are both proportional to the operator $\ket{-}\bra{+}$ causing the decay of the population of $\ket{+}$, in both regimes (with a different relaxation rate though). The operator $\hat s_{-,0}$ and the Lamb shift Hamiltonian commute with this stationary state. Once written back into the lab frame, the state $\ket{-}$ becomes the stationary orbit:
\bb
\ket{\Psi_-(t)} = U_\text{rot}^\dagger U_I^\dagger \ket{-},
\ee
given in Eq.~\eqref{eq:orbit} up to a global phase.

\section{Definition of work exchanged with the bath and associated second law}\label{app:2ndLaw}

In this appendix, we follow the lines of Ref.~\cite{Elouard22} to justify our definition of work exchanged with the bath, by demonstrating that the associated heat (the rest of the energy variation of the bath) fulfills the Second law of thermodynamics. For this, we focus on preparation 1, but assume that the qubit and the bath are held in contact with an environment at finite inverse temperature $\beta$. This results in the initial factorized state $\rho(0) = \ket{\uparrow}_z\bra{\uparrow}_z\otimes \rho_R(0)$, with
\bb
\rho_R(0) &=& \frac{e^{-\beta \left(H_R+R \right)}}{Z_R(0)},\\
&=& {\cal D}[\chi_0]\tilde\rho_R(0){\cal D}^\dagger[\chi_0],
\ee
where $\tilde\rho_R(0)= \frac{e^{-\beta H_R}}{Z_R(0)}$ and ${\cal D}[\chi_0]$ is a collective displacement operator displacing each mode $b_k$ by an amplitude $-g_k/\omega_k$. At vanishing temperature, i.e. $\beta\to\infty$, we retrieve the output of the preparation (1) considered in Sec.~\ref{sec:multimode}. At time $t$, we  define the collective displacement ${\cal D}[\chi_t]$ which displaces each mode $b_k$ by an amount $-\moy{b_k(t)}$, that is, which brings the final state of the bath at time $t$ to a state $\tilde\rho_R(t) = {\cal D}[\chi_t]^\dagger\rho_R(t){\cal D}[\chi_t]$ whose phase-space distributions are centered around the origin. We now use the following identity, for a transformation between $t=0$ (end of the preparation) and $t$:
\bb
\Delta S_R(t) = -D\left(\tilde\rho_R(t) \| \tilde\rho_R(0) \right) -\beta\text{Tr}\{{\cal H}_R(\tilde \rho_R(0)-\tilde\rho_R(t))\},\label{eq:SRQ}
\ee
where $D(\rho\|\sigma) = \text{Tr}\{\rho(\log\rho-\log\sigma)\}$ is the relative entropy of $\rho$ and $\sigma$, a positive quantity. Eq.~\eqref{eq:SRQ} can be straightforwardly demonstrated by expanding the right-hand side terms and using that $\tilde\rho_R(t)$ is related to $\rho_R(t)$ by a unitary, and has therefore the same von Neumann entropy.
Moreover, as the evolution of the system and the bath is unitary, the total von Neumann entropy $S_\text{tot}$ is conserved. Denoting $S_{S}(t)$ (resp. $S_R(t)$) the von Neumann entropy of the system (resp. the bath), we can write:
\bb
\Delta S_S(t) + \Delta S_R(t) = I_{SR}(t),\label{eq:SSI}
\ee
where $I_{SR}(t) = S_S(t)+S_{R}(t)-S_{tot}(t) \geq 0$ is the mutual information that has built up between the system and the bath at time $t$ (it is zero at $t=0$ due to the factorized structure of the initial state).
Injecting Eq.~\eqref{eq:SRQ} into Eq.~\eqref{eq:SSI}, one obtains:
\bb
\Delta S_S(t) -\beta\text{Tr}\{{\cal H}_R(\tilde \rho_R(0)-\tilde\rho_R(t))\}= I_{SR}(t) + D\left(\tilde\rho_R(t) \| \tilde\rho_R(0) \right) \geq 0.\label{eq:2ndLaw}
\ee
We interpret this last inequality as the (out-of-equilibrium version of the) second law of thermodynamics for the system's dynamics, where the role of the heat exchanged (counted with positive sign when energy is received by the system) is played by:
\bb
Q_R(t) &=& \text{Tr}\{{\cal H}_R(\tilde \rho_R(0)-\tilde\rho_R(t))\},\\
&=& \sum_k \omega_k   \text{Tr}\{ b_k^\dagger b_k \left({\cal D}^\dagger[\chi_\uparrow]\rho_R(0){\cal D}[\chi_\uparrow]- {\cal D}^\dagger[\chi_t]\rho_R(t){\cal D}[\chi_t]\right)\}.
\ee
We now use that ${\cal D}[\chi_t]b_k{\cal D}^\dagger[\chi_t] = b_k-\moy{b_k(t)}$, with $\moy{b_k(0)}=-g_k\omega_k$, to simplify the above equation and obtain:
\bb
Q_R(t) &=& -\Delta \sum_k \omega_k \text{Tr}\{ (b_k^\dagger -\moy{b_k}^*) (b_k^\dagger -\moy{b_k}^*) \rho_R\},\\
&=& - \Delta \sum_k \omega_k \left(\moy{b_k^\dagger b_k}-\vert\moy{b_k}\vert^2\right).\label{eq:QRepxr}
\ee
As in the main text, the $\Delta$ symbol refers to a variation between time $t=0$ and time $t$.

By virtue of the first law of thermodynamics, the work provided by  the bath is then (with the same sign convention):
\bb
W_R(t) &=& - \Delta \moy{{\cal H}_R} - Q_R(t),\nonumber\\
&=&\text{Tr}\{{\cal H}_R(\rho_R(0)-\tilde \rho_R(0)+\rho_R(t)-\tilde\rho_R(t))\},\nonumber\\
&=& -\Delta\sum_k \omega_k  \vert \moy{b_k}\vert^2.\label{eq:WRexpr}
\ee

Recalling the energy balance introduced in Sec.~\ref{sec:energetics} in the main text:
\bb
W_\text{dr} &=& \Delta E_R + \Delta E_\text{int}+\Delta E_S,
\ee
and injecting the expressions for heat and work above gives:
\bb\label{eq:1stlaw}
W_\text{dr} &=& -W_R-Q_R + \Delta E_\text{int}+\Delta E_S.
\ee
One can therefore rewrite Eq.~\eqref{eq:2ndLaw} under the form:
\bb
W_\text{dr}(t) + W_R(t) \geq \Delta F_S + \Delta E_\text{int},
\ee
with $F_S(t) = E_S(t) - \beta^{-1}S_S(t)$ the free energy of the spin at inverse temperature $\beta$. This allows us to treat $W_R$ on an equal footing with $W_\text{dr}$ as work exchanged with the spin to vary its free energy (or the energy stored in the coupling Hamiltonian).\\

We finally mention that other nonequilibrium resources beyond coherent displacements can in principle be consumed in the bath state to extract work (or obtain effects similar to a work expense, e.g. a decrease of the spin entropy). The present treatment could be extended to include (i) squeezing, (ii) ergotropy, and even (iii) any deviation from a thermal state by replacing the displacement operators used to define $\tilde\rho_R$ with (i) squeezing operators, (ii) arbitrary unitaries, and (iii) arbitrary entropy-preserving operations, respectively. These different choices correspond to more and more precise expressions for the second law and estimations of the heat, which are each relevant if one can indeed access/quantify more and more complex resources in the bath (or if these resources are initially present in the bath states). We refer to \cite{Elouard22} for the general case. Here, we consider the practically relevant case where work from the bath can only be extracted via coherent displacement, and treat any other energy variation as heat.\newline

\noindent \emph{Efficiency}\\

In the limit of vanishing initial bath temperature considered in the main text, Eq.~\eqref{eq:2ndLaw} ensures that the heat $Q_R$ provided by the bath is negative. From Eq.~\eqref{eq:1stlaw} see that the efficiency of conversion of work performed by the drive $W_\text{dr}$ into work stored in the reservoir $-W_R$ fulfills:
\bb
\eta_\text{dis} &=& \frac{W_\text{dr}+Q_R-\Delta E_\text{int}-\Delta E_S}{W_\text{dr}},\nonumber\\
&=&1-\frac{\Delta E_\text{int}+\Delta E_S-Q_R}{W_\text{dr}}.
\ee
This quantity is not ensured to be lower than $1$ as the energy initially stored in the coupling Hamiltonian can in principle be consumed and converted into work (although not cyclically). To exclude this contribution from our account and focus on the additional work cyclically extractable from the drive, we define the dynamo energy as:
\bb
E_\text{dyn} = \sum_k \omega_k\vert \moy{b_k}+\frac{g_k}{\omega_k}\moy{\sigma_z}\vert^2,
\ee
which can be related to the efficiency $\eta\leq 1$ defined in Eq.~\eqref{eq:efficiency}.

\section{Numerical analysis of fluctuations in the bath}\label{app:fluctuations}
Here we compare the work $W_\text{dr}$, $\Delta E_\text{spin}$, $\Delta E_\text{dyn}$ and $\Delta E_\text{fluct}$ from numerical results.\\
Note that from the definition of the fluctuation energy defined in Eq.~\eqref{eq:E_fluct}, we can identify different contributions to it:
\bb
E_\text{fluct}(t) &=& \sum_k \omega_k \bigg( \moy{ \left(b_k(t)^\dag + \frac{g_k}{\omega_k}S(t)\right)\left (b_k(t) + \frac{g_k}{\omega_k}S(t)\right)}- \left\lvert \moy{b_k(t)} + \frac{g_k}{\omega_k}\moy{S(t)} \right\rvert^2 \bigg), \nonumber \\
&=& \sum_k \omega_k \Bigg[ \bigg( \moy{ b_k^\dag b_k}- \moy{ b_k(t)^\dag}\moy{ b_k(t)} \bigg) + \frac{g_k}{\omega_k} \bigg( \moy{S(t)(b_k^\dag + b_k)} - \moy{S(t)}\moy{b_k^\dag + b_k} \bigg) \nonumber\\
&+& \frac{g_k^2}{\omega_k^2} \bigg( \moy{S(t)^2}-\moy{S(t)}^2 \bigg)\Bigg], \nonumber \\
&=& \sum_k E_\text{fluct,k}^\text{bath} + E_\text{fluct,k}^\text{int} + E_\text{fluct,k}^\text{spin}.
\ee
In ED, we can directly read out all the quantities and in particular we can also decompose the summation over bath modes for $\Delta E_\text{dyn}$ and $\Delta E_\text{fluct}$. This allows in this case to study all the contributions separately. Moreover, we can also study contributions stemming from each of the discretized modes. Let us first present some results for the case of a spin coupled to only one resonant mode.

\subsection{Results for one resonant mode from ED}\label{app:fluctuations_1ED}

\begin{figure}[hb]
    \centering
    \includegraphics[width=\linewidth]{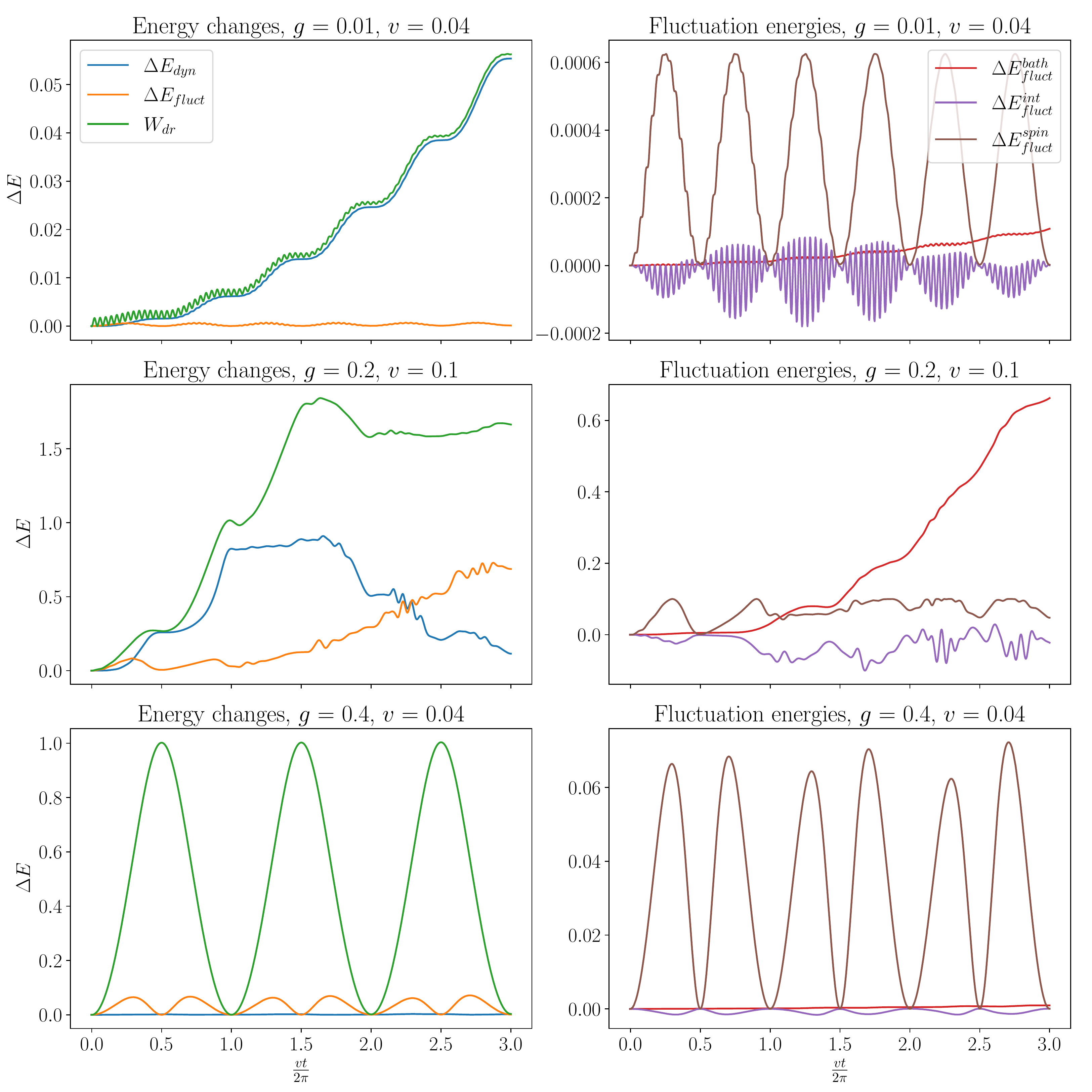}
    \caption{Numerical results from ED for a spin coupled to one resonant mode for different value of the coupling strength $g$ and the velocity $v$. In these simulations, we set $H=1.0$ and we truncate the Hilbert space of the bosonic degree of freedom at a value which we verify to be well above the maximally reached occupation.}
    \label{fig:one_mode_fluctuation}
\end{figure}

Results for the spin coupled to only one resonant mode are presented in Fig.~\ref{fig:one_mode_fluctuation}. We see that in the regime of an efficient dynamo (i.e. when the work done by driving the system is almost entirely consumed by the change in dynamo energy $\Delta E_\text{dyn}$, the latter dominates the change in the fluctuation energy $\Delta E_\text{fluct}$ (upper left plot of Fig. \ref{fig:one_mode_fluctuation}). The largest contribution to the fluctuations comes from spin fluctuations (upper right plot). In the middle plots, we see a regime of stronger coupling in which the initially efficient dynamo breaks down after a few rotations. Consequently, the fluctuation energy of the bath starts to ramp up (middle right plot), so that the total fluctuation energy eventually dominates the dynamo energy. In the lower plots of Fig.~\ref{fig:one_mode_fluctuation}, we see a regime in which the spin is frozen and no dynamo effect occurs, the fluctuation energy thus dominates the dynamo energy, while the work leads merely to excitation of the spin from its ground state, so that the spin-related part of the fluctuation energy is the most relevant one.

\subsection{Results for several modes from ED}\label{app:fluctuations_NED}
\begin{figure}
    \centering
    \includegraphics[width=\linewidth]{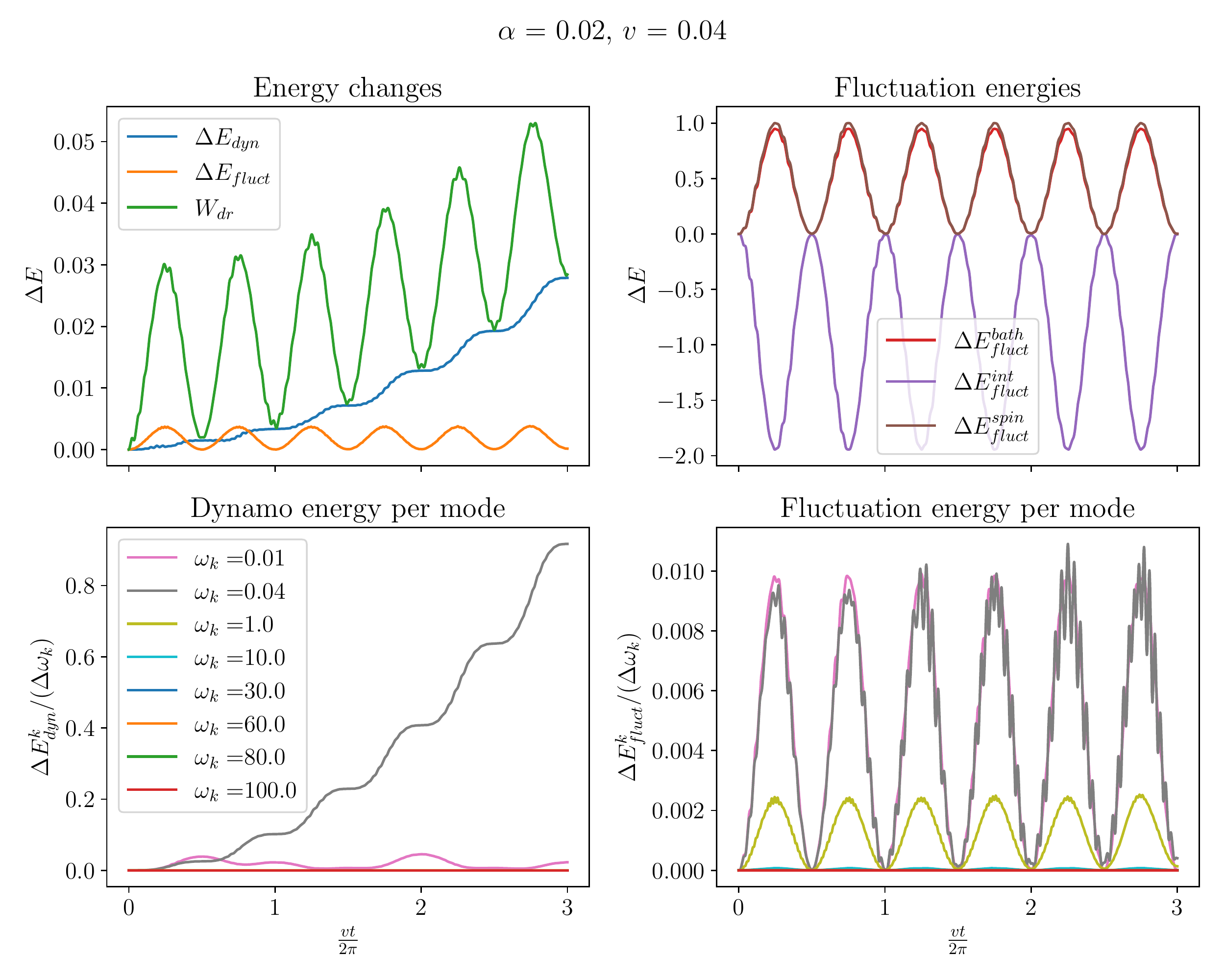}
    \caption{Upper plots: Energy contributions for the total work and the change in dynamo and fluctuation energy, summed over all modes (left). Different contributions to the fluctuation energy (right). Lower plots: Individual contributions to the dynamo energy (left) and the fluctuations (right) coming from each of the simulated modes, normalized by the width of the part of the spectrum discretized by this mode. Here, we used an Ohmic spectral density with a hard cut-off at $\omega_c = 100.0$, discretize it by $N=8$ modes and check that their occupations remain well below the value at which they are truncated. Furthermore, we set $H=1.0$.}
    \label{fig:N_mode_fluctuation}
\end{figure}

To depart from the one mode dynamo, we simulated a spin coupled to a (small) finite number of bosonic modes using ED, as described in the main text in \ref{sec:finite_number_modes}. In this case, we can notably study the contributions of the dynamo energy and the fluctuation energy coming from each mode. Some results are shown in Fig.~\ref{fig:N_mode_fluctuation}. These results are obtained in the regime where an efficient dynamo effect occurs. At the poles, the dynamo energy dominates the fluctuation energy and running the dynamo for some time, it will also start to dominate for all times. We see that the main contribution to the dynamo energy comes from the resonant mode. While the total fluctuation energy is oscillating with an amplitude below the dynamo energy, its individual components can take large values which cancel each other. Note that at the poles, all components vanish. The numerical analysis is limited by the fact that the resonant mode reaches large occupations, which require a larger Hilbert space, making ED simulations difficult. Therefore, for stronger coupling, we resort to the stochastic approach.

\subsection{Results from SSE}\label{app:fluctuations_SSE}
\begin{figure}
    \centering
    \includegraphics[width=\linewidth]{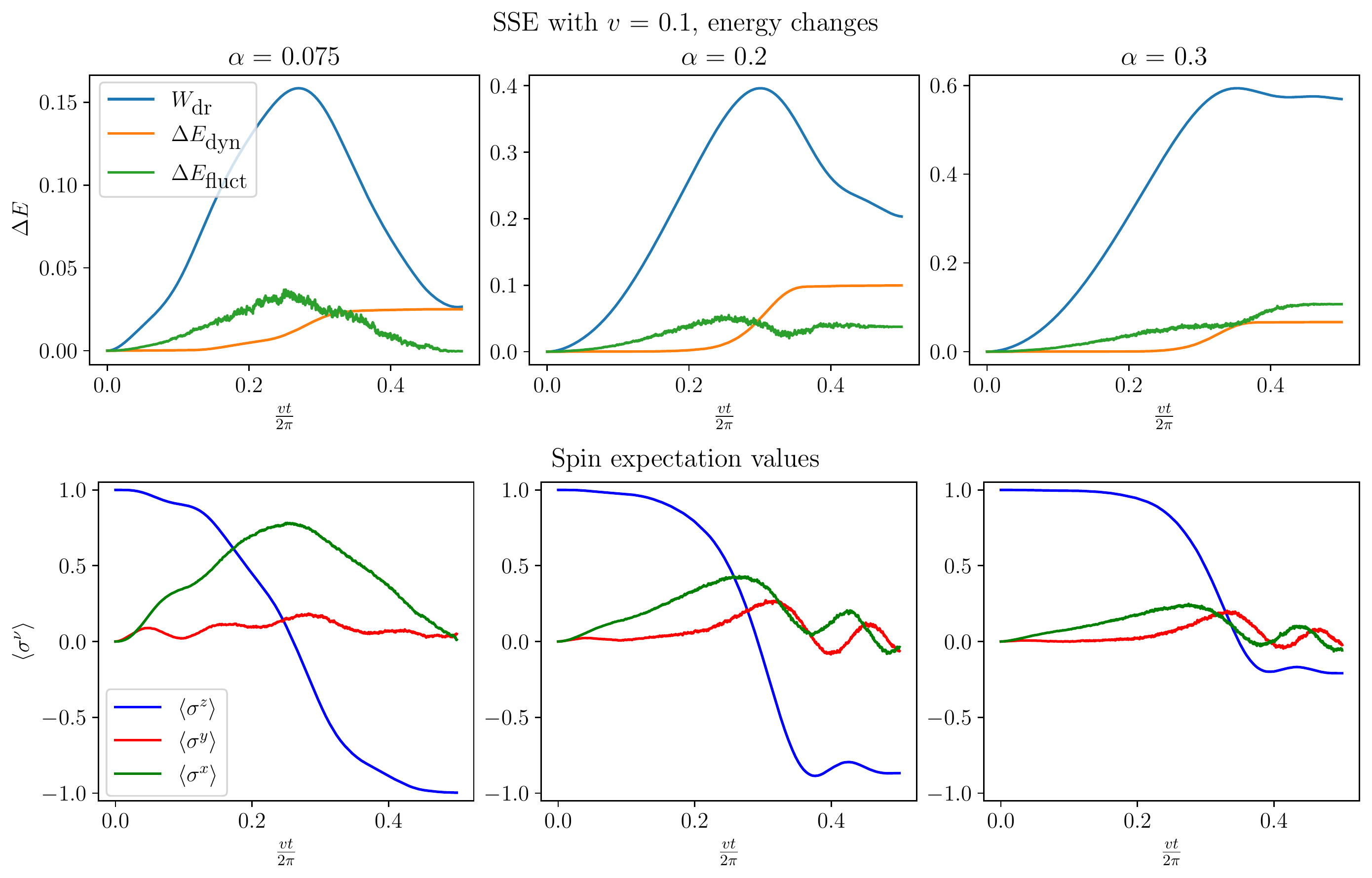}
    \caption{Results from SSE with an Ohmic spectral density with exponential cut-off and $\omega_c = 100.0$, while $H=1.0$. The upper line shows results for the changes in the relevant energies, while the lower line shows the corresponding spin expectation values.}
    \label{fig:SSE_fluctuation}
\end{figure}

Using the results from the stochastic approach, we obtain the spin dynamics and can infer on $W_\text{dr}$, $\Delta E_\text{S}$ and $\Delta E_\text{dyn}$. The fluctuation energy $\Delta E_\text{fluct}$ can then be calculated from the energy balance in Eq.~\eqref{eq:energy_balance}.
Some numerical results for driving during half a period for different values of $\alpha$ are shown in Fig.~\ref{fig:SSE_fluctuation} with a comparison to the corresponding spin expectation values. We see that in the regime of an efficient dynamo and when $C_\text{dyn} \approx 1$ (left column), the change in dynamo energy dominates the fluctuation energy. Consequently, when $C_\text{dyn}$ starts to decrease (middle column), the efficiency of the dynamo decreases (i.e. $\Delta E_\text{dyn}<W_\text{dr}$), until at even stronger coupling (right column) the fluctuations dominate over the change in dynamo energy.
\newpage
\newpage
\newpage

\end{document}